\documentclass[aps,prb,reprint,amsmath,amssymb,nobibnotes]{revtex4-1}

\usepackage{bm}
\usepackage{graphicx}

\begin{document}

\title{Current Fluctuations Driven by Ferromagnetic and Antiferromagnetic Resonance}

\author{Arne Brataas}
\affiliation{Center for Quantum Spintronics, Department of Physics, Norwegian University of Science and Technology, NO-7491 Trondheim, Norway}
\thanks{Arne.Brataas@ntnu.no}

\begin{abstract}
We consider electron transport in ferromagnets or antiferromagnets sandwiched between metals. When spins in the magnetic materials precess, they emit currents into the surrounding conductors.  Generally, adiabatic pumping in mesoscopic systems also enhances current fluctuations. We generalize the description of current fluctuations driven by spin dynamics in three ways using scattering theory. First, our theory describes a general junction with any given electron scattering properties. Second, we consider antiferromagnets as well as ferromagnets. Third, we treat multiterminal devices. Using shot noise-induced current fluctuations to reveal antiferromagnetic resonance appears to be easier than using them to reveal ferromagnetic resonance. The origin of this result is that the associated energies are much higher as compared to the thermal energy. The thermal energy governs the Johnson-Nyquist noise that is independent of the spin dynamics. We give results for various junctions, such as ballistic and disordered contacts. Finally, we discuss experimental consequences.
\end{abstract}
\date{\today}
\maketitle

\section{Introduction}
\label{intro}

In conductors, a bias voltage generates a net current.  However, the current also fluctuates. Noise exists even at equilibrium when the bias voltage is zero. At equilibrium, the current fluctuations are related to the conductance via the fluctuation-dissipation theorem as Johnson-Nyquist noise \cite{Johnson:PR1928,Nyquist:PR1928}. When there are non-zero bias voltages comparable to or larger than the thermal energy, the fluctuation-dissipation theorem does not apply. Instead, the current fluctuates due to shot noise since the electron flow is in discrete quanta of the elementary charge $-e$. The shot noise reveals quantum transport features in nanostructures \cite{Blanter:PhysRep2000}. 

In electron transport, at low temperatures, the transmission probabilities of waveguide eigenmodes $\{ T_n\}$ determine the shot noise of phase-coherent conductors biased by a voltage $V$:
\begin{equation}
S = \frac{2e^2}{h} e V \sum_n T_n \left( 1 - T_n \right) \, . 
\label{biasshotnoise}
\end{equation}
The shot noise expression \eqref{biasshotnoise} is general and captures the nature of many contacts, such as diffusive, ballistic, and tunnel junctions. The factor $1-T_n$ arises from the Pauli exclusion principle; two electrons cannot simultaneously occupy the same waveguide mode. The sum is over the waveguide modes labeled by $n$. 

Ferromagnets have intriguing transport properties caused by the coupling between electric currents, electron spin currents, and localized spin dynamics. Currents can induce spin dynamics by spin-transfer torques\cite{Berger:PRB1996,Slonczewski:JMMM1996,Tsoi:PRL1998,Myers:Science1999,Katine:PRL2000,Kiselev:Nature2003,Krivorotov:Science2005} and spin-orbit torques\cite{Manchon:PRB2008,Miron:NatMat2010,Miron:Nature2011,Liu:Science2012,Garello:NatNan2013,Mellnik:Nature2014,MacNeill:NatPhys2017,Zhu:PRL2019}. The magnetization direction can be switched, or magnetic oscillations can be induced. These phenomena are of a fundamental importance and might be utilized in magnetic random access memories, spin-torque oscillators or spin-logic devices. These developments have been reviewed in Refs.\ \onlinecite{Silva:JMMM2008,Ralph:JMMM2008,Brataas:NatMat2012,Manchon:RMP2019}. The phenomenon reciprocal to spin-transfer torque is spin pumping, the emission of spin currents into metals induced by spin excitations in adjacent magnets\cite{Monod:PRL1972,Silsbee:PRB1979,Urban:PRL2001,Mizukami:Jpn2001,Tserkovnyak:PRL2001,Heinrich:PRL2003,Mosendz:PRB2010}. Spin pumping exposes details of the transport properties and spin dynamics.

Recently, antiferromagnetic spintronics has attracted considerable interest because of the intrinsic high frequencies, new features in spin dynamics, and robustness with respect to external magnetic field disturbances \cite{Jungwirth:NatNano2016,Bodnar:NatCom2018,Lebrun:Nature2018,Baltz:RMP2018,Gomonay:NatPhys2018,Li:Nature2020,Vaidya:Science2020}. Many of the phenomena in ferromagnets have similar or richer behavior in antiferromagnets. For instance, currents can switch the spin configurations \cite{Wadley:Science2016,Bodnar:NatCom2018,Cheng:PRL2020}, and antiferromagnetic resonance excitations can pump spin currents \cite{Cheng:PRL2014,Kamra:PRL2017,Johansen:PRB2017,Li:Nature2020,Vaidya:Science2020}.

Usually, bias voltages induce electric currents and shot noise as in Eq.\ \eqref{biasshotnoise}. However, out-of-equilibrium currents can be sustained by other methods using temporal external or internal drivers that modify the conductor properties. Oscillating electric and magnetic fields can induce net currents. Such drivers also enhance the electric current noise.  In magnetic systems, dynamical spin excitations produce spin currents \cite{Monod:PRL1972,Silsbee:PRB1979,Urban:PRL2001,Mizukami:Jpn2001,Tserkovnyak:PRL2001,Heinrich:PRL2003,Mosendz:PRB2010,Cheng:PRL2014,Kamra:PRL2017,Johansen:PRB2017,Li:Nature2020,Vaidya:Science2020}. 

Spin pumping also causes additional magnetization dissipation \cite{Tserkovnyak:PRL2001,Brataas:PRL2008,Starikov:PRL2010,Brataas:PRB2011,Liu:PRB2011}. Through the fluctuation-dissipation theorem, this implies that fluctuating spin currents associated with spin pumping and spin transfer as well exist \cite{Foros:PRL2005}. In a recent study, Ref.\  \onlinecite{Ludwig:PRR2020} obtained an expression for the electric (charge) current noise caused by ferromagnetic resonance excitations in ferromagnetic-normal metal-ferromagnetic double tunnel barrier systems.  The particularly nice feature is that the mechanism does not require spin-orbit-induced spin-to-charge conversion such as the spin Hall and inverse spin Hall effects. It is also a new channel for detecting and characterizing ferromagnetic resonance and electron transport in magnetic conductors.

Theoretically, scattering matrices capture electron transport in nanostructures well \cite{Buttiker:PRL1993}. They can also describe current-induced torques \cite{Waintal:PRB2000,Brataas:PRL2000,Brataas:EPJB2001,Stiles:PRB2002,Rychkov:PRL2009,Hals:EPL2010}.  Scattering matrices also capture effects due to temporal external or internal drivers. To the lowest order in the driver frequency, the pumped current is related to the stationary scattering properties \cite{Brouwer:PRB1998}. This feature considerably simplifies the description of adiabatic pumping such as spin-pumping \cite{Tserkovnyak:PRL2001,Brataas:PRB2002,Tserkovnyak:RMP2005}. In general, scattering properties can also describe the enhanced electric current noise due to periodic drivers. For the case when only one waveguide mode is linked to each reservoir, Ref.\ \onlinecite{Moskalets:PRB2004} obtained an expression for the current noise in terms of the dynamical scattering properties of the device. 

We consider a magnet that is in contact with normal metal leads. Our purpose is to obtain a general expression for how spin excitations in ferromagnetic and antiferromagnetic structures generate the thermal and shot noise of the electric current.  To this end, we generalize the results of Ref.\ \onlinecite{Ludwig:PRR2020} in three ways: 1) the formalism is valid for arbitrary junctions, 2) the theory applies to spin dynamics in ferromagnets and antiferromagnets, and 3) multiterminal devices are treated. In this way, we obtain general results for electric current noise driven by spin dynamics in magnetic materials in arbitrary junctions. We will find that, when spin angular momentum is conserved, the noise vanishes when the magnet is insulating. Our results are therefore most relevant for conducting systems.

In antiferromagnets, the deduced expressions for the noise are entirely new to the best of our knowledge. Our general results in the case of ferromagnetic excitations are also new. In the limited case of two-terminal double-barrier tunnel ferromagnetic junctions, our general results agree with the results of Ref.\ \onlinecite{Ludwig:PRR2020} by taking into account random disorder in our formulation. Since we use an entirely different approach, this agreement establishes the consistency of both treatments in this limit. We discuss the fact that other junctions have different behaviors in ferromagnets. 

We have organized the presentation as follows. Our paper first gives the main results and consequences before proceeding section by section with more details of the derivations. The next section \ref{model} introduces the model and presents the main results. We will find that four factors determine the shot noise:  i) The electron-transport-related shot noise coefficients, ii) the driver frequency, iii) the thermal energy, and iv) the spin-dynamics factor. Section \ref{spindynamics} discusses the specifics of the influence of ferromagnetic and antiferromagnetic dynamics driven by magnetic fields that govern the spin-dynamics factor. Then, in section \ref{junctions}, we discuss the shot noise coefficients in various junctions, such as ballistic and disordered contacts, both in antiferromagnets and ferromagnets. We present the general theory of adiabatic pumping-induced electric current noise in section \ref{theory}. Section \ref{spinnoise} applies the general theory in section \ref{theory} to derive the spin dynamics-driven shot noise in section \ref{spindynamics}. We conclude our presentation in section \ref{conclusion}. Finally, we derive the general scattering theory of adiabatic driven enhanced electric current noise in Appendix \ref{adiatheory}.

\section{Model and Main Results}
\label{model}

We consider a magnet embedded between metals (or semiconductors) in an open circuit. At equilibrium, electric currents fluctuate in the metals. In the magnet, there can be thermally induced spin fluctuations or coherent spin precessions caused by external forces. We consider the latter case that the spin dynamics is coherent and dominated by external drivers as in ferromagnetic resonance or antiferromagnetic resonance. It is straightforward to generalize our results to explicitly include contributions from incoherent spin dynamics relevant when the external drive is weak or absent.

\begin{figure}[htp]
\includegraphics[width=0.48\textwidth]{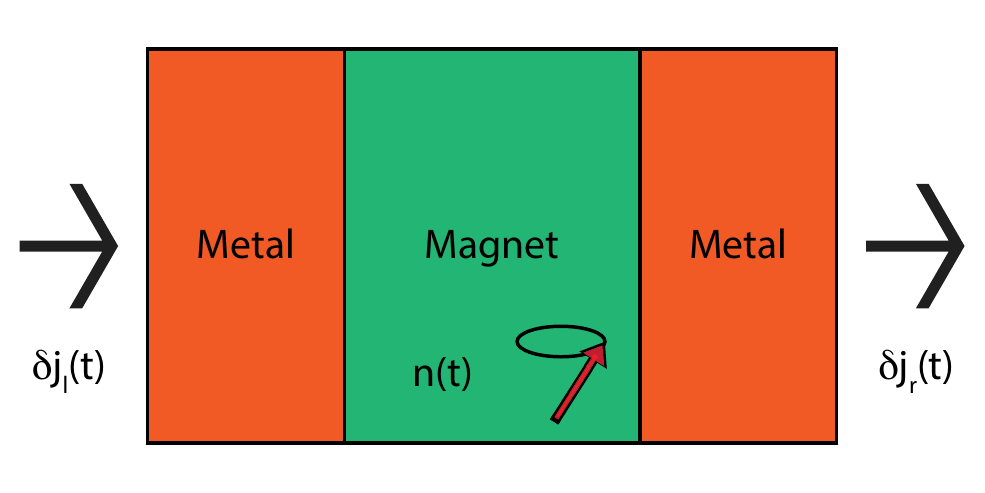}
\caption{Schematic description of a metal-magnet-metal system. An open circuit (not shown) is connected to the system. The electric currents fluctuate. The spin dynamics in the magnet described by precession of the temporal unit vector along the order parameter ${\bf n}(t)$ enhance the current fluctuations.}
\label{noiseres}
\end{figure}

When the spins in the magnet precess, the fluctuations are enhanced. Fig.\ \ref{noiseres} schematically depicts the system in a two-terminal configuration. Our results are also valid for many terminals.
 The unit vector aligned with the order parameter, the magnetization in ferromagnets and the staggered field in antiferromagnets, ${\bf n}$, is homogeneous. When external magnetic fields or currents drive the system, the order parameter ${\bf n}$ precesses around an equilibrium direction. While we subsequently develop a formulation describing general junctions that may include the spin-orbit interaction and magnetic impurity scattering, our first and primary focus is on systems with the conservation of spin angular momentum. Giant magnetoresistance, tunnel magnetoresistance, spin-transfer torques, and spin pumping are examples of central phenomena in such systems.

In systems with the conservation of spin angular momentum, two independent scattering matrices, $S^\uparrow$ and $S^\downarrow$, for spin-up and spin-down electrons govern electron transport. The scattering matrices contain all details of the junctions related to the interfaces between the metals and magnets, band structure, and bulk and surface impurity scattering. We evaluate the electric current and the associated noise in the metallic leads. The electric current direction is towards the magnet. While our formalism is valid irrespective of the magnet`s conducting properties, it is most relevant for metallic systems since we will demonstrate that, in the absence of spin-orbit coupling, the electric noise vanishes when there is no flow of electric charge between the leads. Furthermore, we consider systems where itinerant electrons carry the current and spin currents carried by localized spins can be disregarded. When the spin angular momentum is conserved, our main result is that the low-frequency electric current noise in the presence of coherent spin excitations has two contributions:
\begin{equation}
p_{\zeta \eta}=p^{(\text{th})}_{\zeta \eta}+p^{(\text{sn})}_{\zeta \eta} \, ,
\label{noise}
\end{equation}
where $\zeta$ and $\eta$ label the leads. Electric current conservation ensures that $\sum_\zeta p_{\zeta \eta}=0=\sum_\eta p_{\zeta \eta}$.

In Eq.\ \eqref{noise}, the first term describes thermal Johnson-Nyquist noise, which is independent of the spin dynamics and determined by the conductance tensor $G$ and the thermal energy $k_B T$:
\begin{equation}
p^{(\text{th})}_{\zeta \eta} = (G_{\zeta \eta} + G_{\eta \zeta}) k_B T \, . 
\label{thnoise}
\end{equation}
Microscopically, the conductance tensor is a sum over the scattering properties of two spin components: 
\begin{align}
G_{\zeta \eta} = & \frac{e^2}{h} \text{Tr}_\text{o} \left[\delta_{\zeta \eta} - S^{\uparrow\dag}_{\eta \zeta} S^{\uparrow}_{\zeta \eta}\right] + \nonumber \\
& \frac{e^2}{h} \text{Tr}_\text{o} \left[\delta_{\zeta \eta} - S^{\downarrow\dag}_{\eta \zeta} S^{\downarrow}_{\zeta \eta}\right] \, . 
\label{Gtensor}
\end{align}
The trace, $\text{Tr}_\text{o}$, is over orbital ("o") degrees of freedom only, a sum over the waveguide modes in the leads.

The second and more interesting contribution to the noise in Eq.\ \eqref{noise} is the shot noise driven by the coherent spin dynamics. We obtain our main result for the zero-frequency shot noise:
\begin{equation}
p^{(\text{sn})}_{\zeta \eta} = \frac{A_ {\zeta \eta} +A_{\eta \zeta}}{8} \left[\hbar \omega \coth{\frac{\hbar \omega}{2 k_B T}} - 2 k_B T \right] D(\omega) \, ,
\label{shotnoise}
\end{equation}
where the spin-dynamics control the spin-dynamics factor $D(\omega)$ that is independent of the electron transport properties. We discuss $D(\omega)$ further below. $D(\omega)$ is a positive definite quantity. $\omega$ is the frequency of the spin excitations. The diagonal components of the shot noise of Eq.\ \eqref{shotnoise} are positive definite quantities as are the diagonal components of the shot noise coefficients $A$. The shot noise coefficients $A_{\zeta \eta}$ depend on the electron transport via products of the scattering matrices of spin-up and spin-down electrons:
\begin{align}
A_{\zeta \eta} = & \frac{e^2}{h} \text{Tr}_\text{o} \left [\delta_{\zeta \eta} - \sum_{\alpha \beta}S^\uparrow_{\zeta \alpha} S^{\downarrow \dag}_{\alpha \eta} S^\downarrow_{\eta \beta} S^{\uparrow \dag}_{\beta \zeta}  \right]  + \nonumber \\
& 
\frac{e^2}{h} \text{Tr}_\text{o} \left [\delta_{\zeta \eta} - \sum_{\alpha \beta}S^\downarrow_{\zeta \alpha} S^{\uparrow \dag}_{\alpha \eta} S^\uparrow_{\eta \beta} S^{\downarrow \dag}_{\beta \zeta}  \right] \, .
\label{snpar}
\end{align}
In the case of a two-terminal device, as in Fig.\ \ref{noiseres}:
\begin{equation}
    A_{ll} = \frac{2e^2}{h} \text{Tr}_\text{o} \left[ 1 - (r^\uparrow_{ll} r_{ll}^{\downarrow \dag}+t^\uparrow_{lt} t_{rl}^{\downarrow \dag}) (r^\downarrow_{ll} r_{ll}^{\uparrow \dag}+t^\downarrow_{lt} t_{rl}^{\uparrow \dag}) \right] \, , 
    \label{snpar2}
\end{equation}
where $l$ means left and $r$ means right, $r$ is a reflection coefficient matrix, and $t$ is a transmission coefficient matrix.

In general, the shot noise parameter $A$ of Eq.\ \eqref{snpar} differs from the conductance $G$ of Eq.\ \eqref{Gtensor}, as does the voltage-biased shot noise of Eq.\ \eqref{shotnoise} compared to the average current governed by the conductance $G$. As is well known for the latter case\cite{Blanter:PhysRep2000}, signatures of the junctions and conductors can be distinguished by the ratio between the voltage-biased shot noise parameter and the conductance via the so-called Fano factor, $F=\sum_n T_n (1-T_n)/\sum_n T_n$. For instance, in tunnel junctions $F=1$, and in diffusive wires $F=1/3$ \cite{Blanter:PhysRep2000}. The spin dynamics-driven shot noise reveals more aspects of the electron transport in spin materials. We will compute the central shot noise parameter of Eq.\ \eqref{snpar2} for ballistic and disordered junctions in ferromagnets and antiferromagnets in Section \ref{junctions}.

Ref. \onlinecite{Moskalets:PRB2004} found that the adiabatic pumping driven enhanced noise was related to the behavior of two particles injected into the system. In agreement with this, the shot noise parameter $A$ of Eq.\ \eqref{snpar} contains products of four scattering matrices describing two-particle processes. The new aspect of the shot noise parameter $A$ of Eq.\ \eqref{snpar} is that two of the scattering matrices relate to spin-up electrons, and two relate to spin-down electrons. In contrast, spin pumping is a one-particle process. The spin-mixing conductance \cite{Brataas:PRL2000,Tserkovnyak:PRL2001,Brataas:EPJB2001,Brataas:PRB2002} is a product of one spin-up scattering matrix and one spin-down scattering matrix, two scattering matrices in total. This is because the pumped spin current has spin along the direction transverse to the magnetization direction, a linear combination of spin-up and spin-down states along the spin quantization axis that is parallel to the order parameter. Similarly, we note that the shot noise coefficients of Eq.\ \eqref{snpar} have combinations of spin-up and spin-down properties related to the same lead. Since spin dynamics produce electric (charge) current noise, a natural interpretation is that the fluctuations arise due to temporal fluctuations of the emissions of spin currents. While the emitted spin currents are instantaneously transverse to the order parameter, they can be reconverted to electric (charge) currents at later times due to the spin-filtering effect in magnetic materials.

We observe that the shot noise parameter $A$ vanishes when no transmission occurs between the left and right reservoirs. This can be seen by letting $t^\uparrow \rightarrow 0$ and $t^\downarrow \rightarrow 0$ in Eq.\ \eqref{snpar2} and using the unitarity of the scattering matrices. This behavior implies that no electric noise will occur in spin dynamics-driven metal-magnetic insulator-metal junctions when spin angular momentum is conserved. In contrast, spin pumping and spin-transfer torques can be as efficient in metal-magnetic insulator bilayers as in metal-magnetic conductor bilayers \cite{Brataas:PhysRep2006}. Beyond the formulation in this section that is based on spin conservation, spin-orbit coupling in heavy metals provides a conversion between charge and spin currents so that even magnetic insulators can become noisy \cite{Bender:PRL2019}. Such small effects are proportional to the square of the small spin Hall angle.

In the expression for the shot noise \eqref{shotnoise}, the spin dynamics solely determine the spin-dynamics factor $D(\omega)$. This quantity is small and related to the power absorbed in resonance experiments \cite{Brataas:PRL2008}, which implies that it can be independently measured. At equilibrium and at sufficiently low temperatures, ${\bf n}(t)={\bf n}_0$. Oscillating transverse magnetic fields at frequency $\omega$, ${\bf H}(t) = {\bf H}_+ \exp{i \omega t} +{\bf H}_- \exp{-i \omega t}$, induce small transverse excitations of the order parameter, $\delta {\bf n}={\bf n}_+ \exp{i \omega t} + {\bf n}_- \exp{-i \omega t}$. In the linear response, the changes in the order parameter and the (external or current-induced) magnetic fields are related by the frequency-dependent spin susceptibility $\chi(\omega)$, a $2 \times 2$ matrix in the basis of the transverse coordinates labeled by $i$, so that $n_{i\pm} = \chi_{ij\pm} H_{j\pm}$. In terms of the spin susceptibilities and the oscillating magnetic fields:
\begin{equation}
D (\omega)= \sum_i n_{i+} n_{i-}= \sum_{ijk} \chi_{ij+}  \chi_{ik-}  H_{j+} H_{k-} \, . 
\label{spinfac}
\end{equation}
The spin susceptibilities $\chi(\omega)$ have peaks at the resonance frequencies, as does $D(\omega)$. In section \ref{spindynamics}, we give generic examples for central classes of anisotropies in ferromagnets and antiferromagnets. In the linear response, the transverse excitations are small. Therefore, the factor $D$ is small. Nevertheless, distinguishing the shot noise from the thermal noise should be possible because the former has a strong dependence on the frequency of the driver, while the latter has no such features. Subtracting the frequency-independent background thermal noise reveals the shot noise.

The shot noise of Eq.\ \eqref{shotnoise} takes a different form depending on the ratio between the energy quantum associated with the time dynamics, $\hbar \omega$, and the thermal energy, $k_B T$. At low temperatures, when $\hbar \omega \gg k_B T$, the shot noise becomes:
\begin{equation}
p^{(\text{sn})}_{\zeta \eta} \approx \frac{A_ {\zeta \eta} + A_{\eta \zeta}}{8} |\hbar \omega| D(\omega)\, .
\label{snlowT}
\end{equation}
The shot noise can be distinguished from direct heating by the different frequency dependence. The low-temperature shot noise of Eq.\ \eqref{snlowT} is linear in the absolute value of the excitation frequency $\omega$ relative to the spin-dynamics factor $D(\omega)$ that can be independently measured. 

We find below that $A \sim 2 G$ in many systems. The ratio between the shot noise of Eq.\ \eqref{shotnoise} and the thermal noise of Eq.\ \eqref{thnoise} at low temperatures is then $p_{\zeta \eta}^\text{sn}/p_{\zeta \eta}^\text{th} \sim |\hbar \omega| D({\omega}) /k_B T$. Since the transverse precession angle is small, typically $D(\omega) \sim 10^{-4}$ at resonance, but the possibly large prefactor $|\hbar \omega| /k_B T$ will increase the ratio between the shot noise and the thermal noise from this value. Stronger external drives can also enhance $D(\omega)$. 

In contrast, at high temperatures, $k_B T \gg \hbar \omega $, the shot noise is smaller. We can expand the shot noise in the small parameter $\hbar \omega$ and obtain:
\begin{equation}
p^{(\text{sn})}_{\zeta \eta} \approx \frac{A_ {\zeta \eta} + A_{\eta \zeta}}{8}  \frac{\left(\hbar \omega\right)^2}{6 k_B T} D(\omega) \, .
\label{snhighT}
\end{equation}
At high temperatures, the shot noise of Eq.\ \eqref{snhighT} is suppressed by a factor $|\hbar \omega|/6 k_B T$ with respect to the low temperature limit of the shot noise of Eq.\ \eqref{snlowT}.

Ferromagnets typically have resonance frequencies less than 100 GHz. These frequencies correspond to a low temperature of less than 1 K. Transport measurements in this temperature range can reveal the low-temperature shot noise \eqref{snlowT}. Such and considerably lower-temperature measurements are standard in the study of the fractional quantum Hall effect and require sophisticated cryogenic instrumentation. At the temperature of liquid helium, approximately 4 K, the ratio between the resonance energy and the thermal energy is approximately 0.2.

The resonance frequencies in antiferromagnets can be one to two orders of magnitude higher than those in ferromagnets. Therefore, detecting the low-temperature limit of the shot noise of Eq.\ \eqref{snlowT} appears to be easier for antiferromagnets. Antiferromagnets can have resonance frequencies in the THz range. We can then expect to observe low-temperature shot noise \eqref{snlowT} at temperatures below 10 K when an antiferromagnet precesses at its resonance frequency. At room temperature, the ratio between the high-temperature shot noise of Eq.\ \eqref{snhighT} and the low-temperature shot noise of Eq.\ \eqref{snlowT} is on the order $2 \times 10^{-4}$. Such corrections are small, but their measurement might be possible since corrections due to, e.g., the spin Hall magnetoresistance (SMR), are of a similar magnitude and routinely probed \cite{Nakayama:PRL2013}.

We conclude that detection of low-temperature shot noise \eqref{snlowT} should be possible in antiferromagnets and, with cryogenic techniques, in ferromagnets. Measurement of the high-temperature shot noise \eqref{snhighT} is possible in both systems.

\section{Spin Dynamics}
\label{spindynamics}

In this section, we will compute the spin-dynamics factor $D(\omega)$ in ferromagnets and antiferromagnets.

Consider a uniaxial ferromagnet with the easy axis along the $z$ direction. The free energy density is
\begin{equation}
f_F= - \frac{M}{2\gamma} \omega_A m_z^2 + \delta f_F\, , 
\label{freeF}
\end{equation}
where ${\bf m}$ is a unit vector along the magnetization with magnitude $M$ and $\omega_A$ is the anisotropy energy. A transverse oscillating magnetic field drives the spin dynamics via the additional contribution to the free energy, $\delta f_F = \omega_{H\perp}  \left( m_x \cos{\omega t} + m_y \sin{\omega t} \right) M/\gamma$, where $\omega_{H\perp} $ is the magnitude of transverse magnetic field in units of frequency. We compute the spin susceptibility that governs the spin dynamics factor \eqref{spinfac} from the Landau-Lifshitz-Gilbert equation
\begin{equation}
\frac{\partial {\bf m}}{\partial t} = - {\bf m} \times \bm{\omega}_\text{eff} + \alpha {\bf m} \times \frac{\partial {\bf m}}{\partial t} \, ,
\end{equation}
where the effective field ${\bm \omega}_\text{eff}$  depends on the free energy density \eqref{freeF} as $\bm{\omega}_\text{eff}=- \gamma \delta f_F/ M\delta {\bf m}$ and $\alpha$ is the Gilbert damping constant. In linear response, the spin dynamics factor \eqref{spinfac} then becomes
\begin{equation}
D_F(\omega) = \frac{\omega_{H_\perp}^2} {2\left[ (\omega-\omega_A)^2 + \alpha^2 \omega^2 \right]} \, .
\label{spinfacF}
\end{equation}
As in Eq.\ \eqref{spinfac}, the spin dynamics factor of Eq.\ \eqref{spinfacF} is quadratic in the transverse fields, represented by their magnitudes $\omega_{H\perp} $ in units of frequency. At resonance, $D_F(\omega_A)= (\omega_{H\perp}/\alpha \omega_A)^2/2$. 

Similarly, we can consider a uniaxial antiferromagnets with the easy axis along the $z$ direction. The free energy density is:
\begin{equation}
f_{AF}= \frac{L}{2 \gamma} \left[ \omega_E {\bf m}^2 - \omega_A n_z^2 \right] + \delta f_{AF}\, , 
\label{freeAF}
\end{equation}
where ${\bf n}$ is a unit vector along the staggered field, ${\bf m}$ is the dimensionless small magnetic moment, $L$ is the magnitude of the staggered magnetization, $\gamma$ is the gyromagnetic ratio, $\omega_E$ is the exchange energy, and $\omega_A$ is the anisotropy energy. A transverse oscillating magnetic field drives the spin dynamics via the additional contribution to the free energy, $\delta f_{AF} = \omega_{H\perp}  \left( m_x \cos{\omega t} + m_y \sin{\omega t} \right) L/\gamma$. The coupled Landau-Lifshitz-Gilbert equations for the staggered field {\bf n} and the magnetic moment {\bf m} are 
\begin{align}
\frac{\partial {\bf n}}{\partial t} = & - {\bf n} \times \bm{\omega}_\text{m,eff} -  {\bf m} \times \bm{\omega}_\text{n,eff} \nonumber \\
& + \alpha {\bf n} \times \frac{\partial {\bf m}}{\partial t} + \alpha {\bf m} \times \frac{\partial {\bf n }}{\partial t} \, , \\
\frac{\partial {\bf m}}{\partial t}  = &- {\bf n} \times \bm{\omega}_\text{n,eff} -  {\bf m} \times \bm{\omega}_\text{m,eff} \nonumber \\
& + \alpha {\bf n} \times \frac{\partial {\bf n}}{\partial t} + \alpha {\bf m} \times \frac{\partial {\bf m }}{\partial t} \, ,
\end{align}
where the effective fields $\bm{\omega}_\text{n,eff}$ and $\bm{\omega}_\text{m,eff}$ depend on the free energy density \eqref{freeAF} as $\bm{\omega}_\text{n,eff}= - \gamma \delta f_{AF}/ L \delta {\bm n}$ and $\bm{\omega}_\text{m,eff}= - \gamma \delta f_{AF}/ L \delta {\bm m}$.

In linear response, and in the exchange approximation, $\omega_E \gg \omega_A$, the spin dynamics factor \eqref{spinfac} becomes:
\begin{equation}
D_{AFM}(\omega) = \frac{\omega^2 \omega_{H_\perp}^2}{2 (\omega^2 - \omega_r^2)^2 + 8 \alpha^2 \omega^2 \omega_E^2} \, ,
\label{spinfacAF}
\end{equation}
where $\omega_r=\sqrt{2 \omega_A \omega_E}$ is the resonance energy and $\alpha$ is the Gilbert damping constant. As in Eqs.\ \eqref{spinfac} and \eqref{spinfacF}, the spin dynamics factor of Eq.\ \eqref{spinfacAF} is quadratic in the transverse fields, represented by their magnitudes $\omega_{H\perp} $ in units of frequency. At resonance, $D_{AFM}(\omega_r)=(\omega_{H\perp}/\alpha \omega_E)^2/8$.

Generalizations to other anisotropies and the inclusion of effects arising from external magnetic fields are straightforward in both antiferromagnets and ferromagnets.

\section{Junctions}
\label{junctions}

In this section, we compute the shot noise coefficients $A$ for simple models of ballistic and disordered junctions. Beyond the scope of the present paper, extensions of these calculations are feasible. Generalizations to consider the effects of the band structure with ab initio calculations and more complicated models of junctions and disorder are possible. Similar calculations have been successfully carried out for interface resistances \cite{Schep:PRB1997}, spin-transfer torques \cite{Xia:PRB2002}, spin pumping \cite{Zwierzycki:PRB2005}, and Gilbert damping \cite{Starikov:PRL2010,Liu:PRB2011}.

In ferromagnets, the potential landscapes for spin-up and spin-down electrons strongly differ. Therefore, the reflection and transmission amplitudes as well as probabilities are spin dependent. In antiferromagnets, the reflection and transmission probabilities are the same for spin-up and spin-down electrons under compensation of the localized spins. Nevertheless, the quantum mechanical phases associated with reflection and transmission differ for the two spin directions.

\subsection{Clean metal}

In clean, ballistic systems, the waveguide modes experience either perfect transmission or perfect reflection. In a simple semiclassical model of a normal metal-ferromagnet-normal metal junction, we can assume $N_\uparrow$ propagating channels for spin-up electrons and $N_\downarrow$ propagating channels for spin-down electrons. Then,
\begin{equation}
A_{ll} =  \frac{2e^2}{h} P N \, , 
\end{equation}
where $P=(N_\uparrow - N_\downarrow)/(N_\uparrow + N_\downarrow)$ is the polarization and $N=N_\uparrow + N_\downarrow$ is the total number of conducting channels. Similarly, the two-terminal conductance becomes $G_{ll}=e^2 N/h$ so that the ratio between the shot-noise coefficient and conductance is $A_{ll}/G_{ll}=2 P$.

In a similar model of compensated antiferromagnets, $N_\uparrow=N_\downarrow$, and thus,
\begin{equation}
A_{ll}=0 
\end{equation}
while $G_{ll}= e^2 N/h$ so that $A_{ll}/G_{ll}=0$.

Therefore, for this simple semiclassical model, the shot noise vanishes in antiferromagnets. However, this is generically not the case for other kinds of junctions. More realistic models of clean junctions will probably result in a small but finite shot noise coefficient in antiferromagnets as well.

The semiclassical results for clean junctions illustrate that the shot noise coefficients can strongly differ in antiferromagnets and ferromagnets.

\subsection{Disordered Metals}

When sufficient disorder exists, either because of bulk impurity scattering or scattering at boundaries, we can use random matrix theory to evaluate the average of scattering matrices. In the semiclassical regime, the phases of the reflection and transmission coefficients are random. They are also statistically independent for spin-up and spin-down electrons. The averages of the transmission and reflection probabilities are\cite{Schep:PRB1997}:
\begin{equation}
T_{ij}^\sigma= \frac{1}{N} \frac{1}{1+\pi_\sigma}
\label{TRMT}
\end{equation}
and
\begin{equation}
R_{ij}^\sigma =  \frac{1}{N} \frac{\pi_\sigma}{1+\pi_\sigma} \, ,
\label{RRMT}
\end{equation}
where $\pi^\sigma=\rho_\sigma d N e^2/A h$, $N$ is the number of waveguide modes, $\rho_\sigma$ is the resistivity for each spin direction, $d$ is the width of the junction, $A$ is the cross section of the junction, and the spin directions are $\sigma=\uparrow$ and $\sigma=\downarrow$. We can then compute that the spin-dependent conductance is $G_\sigma= (e^2/h) \sum_{ij} T_{ij} = G_{d\sigma}/(1+G_{d\sigma}/G_{sh})$, where the conductance of a diffusive conductor is $G_{d\sigma}= A/ \rho_\sigma d$ and the Sharvin conductance is $G_{sh}=e^2 N/h$. A more intuitive expression is that the resistance consists of the Sharvin resistance in series with the diffusive resistance, $1/G_\sigma = 1/G_{sh} + 1/G_{d\sigma}$. The total conductance is $G=G_\uparrow + G_\downarrow$.

In ferromagnets, the conductances for spin-up and spin-down electrons differ. Based on Eqs.\ \eqref{TRMT} and \eqref{RRMT}, we can now obtain the average:
\begin{align}
\langle A_{ll} \rangle & =  2 \left(G_\uparrow + G_\downarrow - 2 \frac{G_\uparrow G_\downarrow}{G_{sh}} \right) \, . 
\end{align}
In the diffusive regime, $G_\uparrow, G_\downarrow \ll G_{sh}$, and we obtain $A_{ll}= 2 G$. This result agrees with the results computed in Ref.\ \onlinecite{Ludwig:PRR2020} for a double barrier tunnel junction system. While the transport regimes in our approach and Ref.\ \onlinecite{Ludwig:PRR2020} are not identical, the treatments seem to share the common feature that strong randomization of the electron trajectories occurs. It is, therefore, natural that the results agree in this limited case.

In compensated antiferromagnets, spin-up and spin-down electrons have the same conductance, $G^\uparrow=G^\downarrow=G/2$. However, the phases of the reflection and transmission coefficients for the spin-up and spin-down electrons remain statistically independent, as in ferromagnets. Then, the shot noise coefficient is:
\begin{align}
\langle A_{ll} \rangle & =  2\frac{e^2}{h} \left(G -  \frac{G^2}{2G_{sh}} \right) \, , 
\end{align}
and in the diffusive limit, we obtain the same result as for a ferromagnet, $A_{ll}= 2G$.

We conclude that for disordered ferromagnets and antiferromagnets, the ratio between the shot noise coefficient and the two-terminal conductance is $A_{ll}/G_{ll}=2$.

\section{Theory of Pumping-Induced Noise}
\label{theory}

We will, in this section, present our general results for noise enhancements by adiabatic pumping. We derive these results from the general scattering theory with multi-terminals and an arbitrary number of waveguide modes in Appendix \ref{adiatheory}. The results in this section are valid for any periodic drive and are not limited to spin-dynamics driven noise discussed in section \ref{model}. We will, in the next section \ref{spinnoise}, use the results in this section to obtain the results for the spin-dynamics drive noise that we presented in section \ref{model}.

We consider phase-coherent conductors attached to reservoirs via leads. Within the conductors, scattering by spin-conserving impurities, the spin-orbit interaction, and the exchange field arising from localized spins can occur. Above, in section \ref{model}, we have assumed that spin angular momentum is conserved and that the magnetization in ferromagnets or staggered fields in antiferromagnets is homogeneous. However, we do not use these assumptions here when presenting the general formula for pumping-induced noise. Appendix \ref{adiatheory} gives details of the derivation of the formulas presented in this section.

\begin{figure}[htp]
\includegraphics[width=0.48\textwidth]{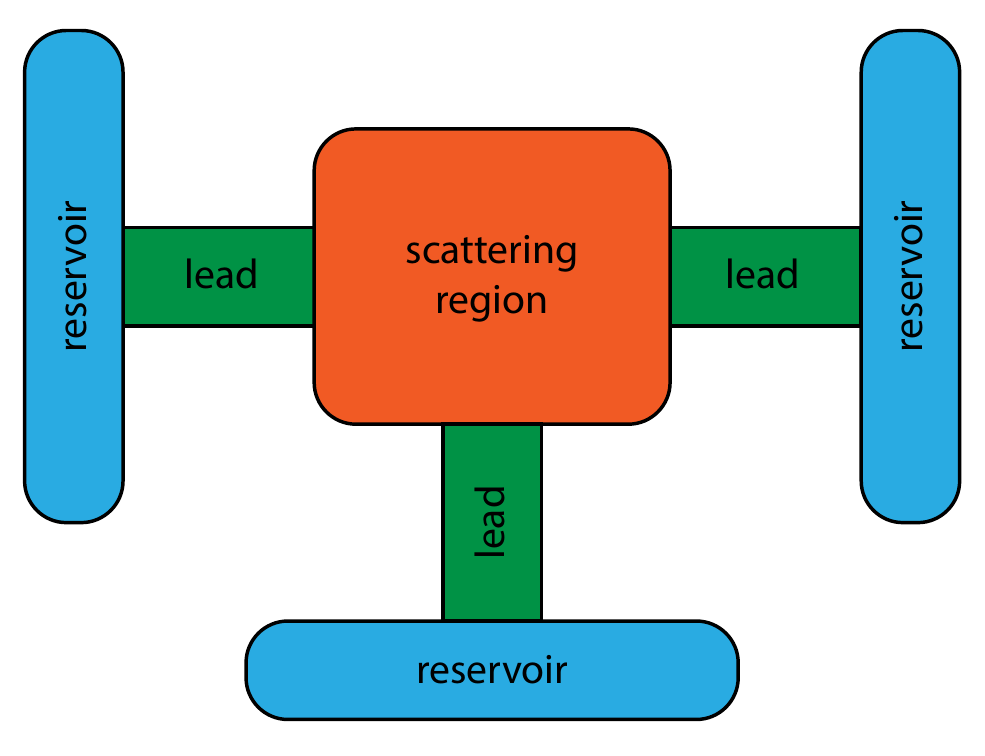}
\caption{Schematic example of a three-terminal device. A scattering region (red area) is connected via leads (green areas) to particle reservoirs (blue areas). Currents can flow between the reservoirs.}
\label{scatreg}
\end{figure}

While we consider a general setup with many reservoirs, we give an example of a three-terminal device in Fig.\ \ref{scatreg}.
Currents can flow between the reservoirs, arising from either differences in bias voltages therein or time-dependent changes within the scattering region. Above, we have considered the latter case when spin excitations drive the scattering region. Our focus is on the current fluctuations when all of the reservoirs are at equilibrium.

We consider a general phase-coherent conductor. Scattering matrices then describe transport between the reservoirs. All orbital waveguide modes and spin quantum numbers span these scattering matrices. In general, the matrices have diagonal and off-diagonal components in orbit and spin. In our case, since the scattering region changes in time, the scattering matrices also have a complex temporal dependence. However, when the temporal changes are slow compared to the typical electron transport time, knowing the temporal behavior of the frozen scattering matrix is sufficient (see Appendix \ref{adiatheory}). We evaluate the frozen scattering matrix at a snapshot in time when the driver has a constant value. This scattering matrix is $S_{\alpha n \gamma l}(t,\epsilon)$, where $\alpha$ is the outgoing lead, $n$ is the outgoing waveguide mode (including spin), $\gamma$ is the incoming lead, $l$ is the incoming waveguide mode (including spin), $t$ is the time, and $\epsilon$ is the electron energy.

The current fluctuations are
\begin{equation}
P_{\zeta \eta}(t_1,t_2) = \frac{1}{2} \langle \Delta I_{\zeta}(t_1) \Delta I_{\eta}(t_2) + \Delta I_{\eta}(t_2) \Delta I_{\zeta}(t_1)\rangle \, , 
\label{curfluct}
\end{equation}
where $\Delta I_{\zeta}(t)=I_\zeta(t)-\langle I_\zeta \rangle(t)$ is the deviation of the current $I_\zeta(t)$ from its expectation value $ \langle I_\zeta(t) \rangle$ in lead $\zeta$. The period of the driver is $T=2\pi/\omega$. Following Ref.\ \onlinecite{Moskalets:PRB2004}, apart from a factor of 2, we define the zero frequency noise as:
\begin{equation}
p_{\zeta \eta} = \int_0^T \frac{1}{T} \int_{-\infty}^\infty d\tau P_{\zeta \eta}(t+\tau/2,t-\tau/2) \, .
\label{znoise}
\end{equation}
Our first central step is that we compute a general expression for the noise induced by a slowly and periodic varying change in the scattering region. When the elastic transport properties are weakly energy dependent,
the current cross correlations are:
\begin{align}
p_{\zeta \eta} & =\sum_q X_{\zeta \eta}^{(s)}(\hbar \omega_q) k_B T \nonumber \\
& +\frac{1}{2} \sum_{q} X_{\zeta \eta}^{(s)}(\hbar \omega_q) \left[ \hbar \omega_q \coth{\frac{\hbar \omega_q}{2 k_B T}} - 2k_B T \right] \, , 
\label{crosscorr}
\end{align}
where the first term represents the thermal noise contribution and the second represents the shot noise contribution. The frequency quantum $\hbar \omega_q$ relates to the period $T$ of the driver by $\hbar \omega_q = 2 \pi q/T$, where $q$ is an integral number. The coefficients $X_{\zeta \eta}^{(s)}(\hbar \omega)= [X_{\zeta \eta}(\hbar \omega) +X_{\zeta \eta}(-\omega_q)]/2$ are determined by the scattering matrices:
\begin{equation}
X_{\zeta \eta}(\hbar \omega_q) = \frac{e^2}{h}\sum_{n_\zeta n_\eta} \sum_{\beta m \gamma l} \Phi_{\zeta n_\zeta \beta m \gamma l}(\omega_q) \Phi_{\eta n_\eta \gamma l \beta m}(-\omega_q) \, ,
\label{X}
\end{equation}
where
\begin{equation}
\Phi_{\zeta n_\zeta \beta m \gamma l}(\omega_q) =\frac{1}{T} \int_0^T dt e^{-i\omega_q t} \Phi_{\zeta n_\zeta \beta m \gamma l}(t) \, ,
\label{Phiq}
\end{equation}
\begin{equation}
\Phi_{\alpha n \beta m \gamma l}(t) = 
\left [ \delta_{\alpha n \beta m}  \delta_{\alpha n \gamma l}   -  S^*_{\alpha n \beta m} (t)
 S_{\alpha n \gamma l}(t)
 \right ] \, , 
 \label{Phit}
 \end{equation}
and the static ("frozen") scattering matrices are to be evaluated at the Fermi energy.

The result for the thermal and shot noise of Eq.\ \eqref{crosscorr} are general for any drivers and valid when the elastic transport properties are weakly energy-dependent. In the next section \ref{spinnoise}, we use this general result to find the noise driven by spin excitations.

\section{Spin Dynamics-Driven Noise}
\label{spinnoise}

In this section, we explain how we can use the general result of the pumping-driven noise in the previous section \ref{theory} to obtain the shot noise when the pumping is due to spin dynamics. We consider homogeneous spin dynamics relevant to ferromagnetic resonance and antiferromagnetic resonance. Now, we assume the conservation of spin angular momentum as in the phenomena of spin-transfer torques and spin pumping. We do not explicitly consider the spin-orbit coupling instrumental relevant for e.g. spin-orbit torques, but further investigations using the same formalism can elucidate its role.

Since the degrees of freedom of the orbital are independent of the spin degrees of freedom, we use the notation that the states $n$ consist of orbital quantum numbers $n_o$ and spin quantum numbers $s$, $n \rightarrow n_o s$. When spin angular momentum is conserved, we separate the frozen S-matrix into spin-independent (labelled by superscript "(c)") and spin-dependent terms (labelled by superscript "(s)"):
\begin{align}
S_{\eta n_o s \zeta m_o s`}   =S^{(c)}_{\eta n_o \zeta m_o}  \delta_{s s`}+ \bm{\sigma}_{ss`} \cdot {\bf n}(t) S^{(s)}_{\eta n_o \zeta m_o} \, , 
\end{align}
where ${\bf n}$ is a unit vector in the direction of the order parameter, ${\bf n}^2=1$. The calculations in this section are valid for both ferromagnets where the order parameter is the magnetization and for antiferromagnets where the order parameter is the staggered field.

As the spins precess, only the spin-dependent part of the scattering matrix acts as a pump. Inserting the spin-dependent scattering matrix into Eq.\ \eqref{X} and using the unitarity of the scattering matrices and the normalization ${\bf n}^2=1$, after considerable algebra, we obtain that the factor that appears in the general expression for the noise of Eq.\ \eqref{crosscorr} becomes:
\begin{align}
 X^{(s)}_{\zeta \eta} & = \int_0^T \frac{dt_1}{T} e^{-i\omega_q t_q} \int_0^T \frac{dt_2}{T} e^{i \omega_q t_2} \times \nonumber \\
 & \left[G_{\zeta \eta}+G_{\eta \zeta} + \frac{A_{\zeta \eta} + A_{\eta \zeta}}{8} \left[ {\bf n}(t_1) - {\bf n}(t_2)  \right]^2 \right] \, ,
 \label{Xspindyn}
 \end{align}
where the conductance tensor $G$ is defined in Eq.\ \eqref{Gtensor} and the shot noise coefficients $A$ are defined in Eq.\ \eqref{snpar}. 

To proceed to find the expression for the spin-dynamics driven shot noise of Eq.\ \eqref{noise} with the thermal contribution of Eq.\ \eqref{thnoise} and the shot noise contribution of Eq.\ \eqref{shotnoise}, we need to evaluate the following integral appearing in the last term of Eq.\ \eqref{Xspindyn}:
\begin{align}
W &= \int \frac{dt_1}{T} e^{-i\omega_q t_q} \int \frac{dt_2}{T} e^{i \omega_q t_2} \left[ {\bf n}(t_1) - {\bf n}(t_2)  \right]^2 
\label{W} 
\end{align}
to the second order in the deviation of the order parameter from equilibrium. To this end, expanding to the linear order is sufficient:
\begin{equation}
{\bf n}(t_1) - {\bf n}(t_2)  = \sum_{\pm} \delta {\bf n}_\pm \left[e^{\pm i \omega t_1} - e^{\pm i \omega t_2} \right] \, , 
\label{orderlinear}
\end{equation}
where $\delta {\bf n}+$ and $\delta {\bf n}_-$ are transverse to the equilibrium spin directions ${\bf n}_0$.

Inserting the linear expansion of Eq.\ \eqref{orderlinear} into Eq.\ \eqref{W}, we then obtain that $W(\hbar \omega_q=0) = 2 \delta {\bf n}_+ \cdot \delta {\bf n}_-$ and $W(\hbar \omega_q=\pm \hbar \omega) = - \delta {\bf n}_+ \cdot \delta {\bf n}_-$. As a consequence, we find
\begin{equation}
X_{\zeta \eta}^{(s)} (\hbar \omega_q=0) = \left( G_{\zeta \eta} + G_{\eta \zeta} \right) + \frac{A_{\zeta \eta} + A_{\eta \zeta}}{8} 2 \delta {\bf n}_+ \cdot \delta {\bf n}_-
\label{Xzero}
\end{equation}
and
\begin{equation}
X_{\zeta \eta}^{(s)} (\hbar \omega_q=\pm \hbar \omega) = - \frac{A_{\zeta \eta} + A_{\eta \zeta}}{8}  \delta {\bf n}_+ \cdot \delta {\bf n}_- \, .
\label{Xfirst}
\end{equation}

For both ferromagnets and antiferromagnetrs, we can now insert the expressions for $X^{(s)}$ of Eqs.\ \eqref{Xzero} and \eqref{Xfirst} into the general expression for the noise of Eq.\ \eqref{crosscorr}. The thermal contribution to the noise is then Eq.\ \eqref{thnoise} and is independent of the spin oscillations. The shot noise contribution is given in Eq.\ \eqref{shotnoise}.

\section{Conclusions}
\label{conclusion}

In conclusion, we have presented general expressions for the noise driven by ferromagnetic and antiferromagnetic resonance. The noise consists of thermal and shot noise contributions. Conductances determine the thermal noise. Shot noise coefficients and the frequency-dependent magnitude of the spin excitations determine the shot noise. The shot noise attains its maximum at ferromagnetic resonance in ferromagnets and antiferromagnetic resonance in antiferromagnets. 

The shot noise parameter can be evaluated for arbitrarily junctions. We have given examples for ballistic systems and disordered systems. The ratio between the spin dynamics-driven shot noise parameter and the conductance is smaller for ballistic systems than for disordered systems. This feature is similar to the behavior of the Fano factor associated with voltage-driven shot noise. 

Our formalism can be generalized to treat the spin-orbit coupling related to spin-orbit torques and electric (charge) pumping \cite{Hals:EPL2010,Ciccarelli:NatNan2015}. Such extensions will shed further light on spin-charge conversions related to spin dynamics.

\begin{acknowledgements}
This work was supported by the Research Council of Norway through its Centres of Excellence funding scheme, project number 262633, "QuSpin". We would like to thank Akashdeep Kamra, Sebastian Goennenwein, and Thomas Tybell for comments on the manuscript.
\end{acknowledgements}

\appendix
\section{Derivation of General Theory of Noise due to Adiabatic Pumping}
\label{adiatheory}

Using Floquet scattering theory, Ref.\ \onlinecite{Moskalets:PRB2004} considered pumping-driven noise in a multiterminal configuration with one-dimensional leads. In other words, each lead only had one waveguide mode. The purpose of the present section is to generalize this description to find equations for arbitrary two- and three-dimensional leads that can also capture the effects of impurities and boundary scattering. To this end, we include many waveguide modes. In our derivation, we also found that an alternative path without explicitly using Floquet scattering states could be easily followed. We will demonstrate that our results agree with the results in Ref.\ \onlinecite{Moskalets:PRB2004} for one-dimensional leads.

In metallic systems, the energy quantum associated with the pump oscillations is typically much smaller than the Fermi energy. In this regime, the relevant previous result is Eq.\ (27) in Ref.\ \onlinecite{Moskalets:PRB2004} when the scattering matrix is weakly energy dependent due to the noise:
\begin{equation}
p_{\text{M}\zeta \eta} =  p_{\zeta \eta}^{(\text{th})} + p_{\text{M}\zeta \eta}^{(\text{sh})} \, .
\label{pM}
\end{equation}
The expression for the thermal noise $p_{\zeta \eta}^{(\text{th})}$ is the same as that in Eq.\ \eqref{thnoise} in the limit of only one mode in all leads. The one-dimensional shot noise contribution in the notation of Ref.\ \onlinecite{Moskalets:PRB2004} is:
\begin{equation}
p_{\text{M}\zeta \eta}^{(\text{sh})} = \frac{2e^2}{h} \sum_{q=1}^\infty   C_{\zeta \eta,q}^{(\text{sym})} \left[ \hbar \omega \coth{\frac{\hbar \omega_q}{2 k_B T}} - 2 k_B T\right]\, , 
\label{pMshotnoise}
\end{equation}
where $C_{\zeta \eta q}^{(\text{sym})} =  [ C_{\zeta \eta q} + C_{\zeta \eta -q}]/2$,
\begin{equation}
C_{\alpha \beta q} = \sum_{\gamma \delta} \left[S^*_{\alpha \gamma} S_{\alpha \delta}  \right]_q \left[S^*_{\beta \delta } S_{\beta \gamma}  \right]_{-q} \, , 
\label{C}
\end{equation}
and the (frozen) scattering matrices should be evaluated at the Fermi energy $E_F$. The Fourier transform of the product of the (frozen) scattering matrices (at the Fermi energy) is defined as follows:
\begin{equation}
\left[S^*_{\alpha \gamma} S_{\alpha \delta}  \right]_q = \int_0^T \frac{dt}{T} e^{i q \omega t} \left[S^*_{\alpha \gamma}(t) S_{\alpha \delta}(t)  \right] \, . 
\label{Mfourier}
\end{equation}
We reproduce the result in Ref.\ \onlinecite{Moskalets:PRB2004} represented by Eqs.\ \eqref{pM}, \eqref{pMshotnoise}, \eqref{C}, and \eqref{Mfourier} for one-dimensional leads and obtain generalizations to leads with an arbitrary number of waveguide modes.

The starting point for our derivation is the expression for the current operator in lead $\alpha$:
\begin{equation}
\hat{I}_\alpha (t) = 2 \pi \hbar e \sum_n   \left[ \hat{a}^\dag_{\alpha n}(t) \hat{a}_{\alpha n}(t)  - \hat{b}^\dag_{\alpha n} (t) \hat{b}_{\alpha n}(t) \right] \, ,
\label{curop}
\end{equation}
where $\alpha$ denotes the lead and $n$ denotes the transverse waveguide mode (orbital and spin). The outgoing operators $\hat{b}$ are related to the incoming operators $\hat{a}$ via the time-dependent scattering matrix $S$:
\begin{equation}
\hat{b}_{\alpha n}(t_1) = \sum_{\beta m} \int_{-\infty}^\infty dt_2 S_{\alpha n \beta m}(t_1,t_2) \hat{a}_{\beta m}(t_2) \, .
\label{curop}
\end{equation}
We use the Fourier transform as follows:
\begin{equation}
\hat{a}_{\beta m} (t) = \frac{1}{2\pi \hbar} \int d\epsilon e^{-i \epsilon t} \hat{a}_{\beta m}(\epsilon)
\label{eq:annihilationFourier}
\end{equation}
and the corresponding inverse Fourier transform.
At thermal equilibrium, the thermal averages are:
\begin{equation}
\langle  \hat{a}^\dag_{\alpha n}(\epsilon_2) \hat{a}_{\beta m}(\epsilon_1) \rangle_\text{eq} = \delta_{\alpha \beta}\delta_{nm} \delta(\epsilon_2-\epsilon_1) f(\epsilon_1)\, ,
\label{eq:FDexpectation}
\end{equation}
where $f(\epsilon)$ is the Fermi-Dirac distribution function that depends on the chemical potential $\mu$ and the thermal energy $k_B T$. The fluctuations are:
\begin{align}
&\langle \hat{a}^\dag_{\alpha k} (\epsilon_1) \hat{a}_{\beta l} (\epsilon_2) \hat{a}^\dag_{\gamma m} (\epsilon_3) \hat{a}_{\delta n} (\epsilon_4)  \rangle - \nonumber \\
&\langle \hat{a}^\dag_{\alpha k} (\epsilon_1) \hat{a}_{\beta l} (\epsilon_2) \rangle \langle \hat{a}^\dag_{\gamma m} (\epsilon_3) \hat{a}_{\delta n} (\epsilon_4)  \rangle \nonumber \\
&= \delta_{\alpha k \delta n} \delta_{\beta l \gamma m} f(\epsilon_1) [ 1 - f(\epsilon_2) ] \delta(\epsilon_1 - \epsilon_4) \delta(\epsilon_2 - \epsilon_3) \, .
\label{eq:fluct}
\end{align}

We express the scattering matrix in terms of the Wigner representation\cite{Rammer:RMP1986}:
\begin{equation}
S(t ,t')=\frac{1}{2\pi \hbar}\int_{-\infty }^{\infty }d\epsilon S\left(%
\frac{t +t'}{2},\epsilon \right)e^{-i\epsilon (t -t')/\hbar } \, .
\label{Wigner}
\end{equation}
The inverse transform is:
\begin{equation}
S(t,\epsilon) = \int_{-\infty}^{\infty} d\tau S(t+\tau/2,t-\tau/2) e^{i \epsilon \tau/\hbar} \, . 
\end{equation}
By Taylor expanding the S-matrix $S((t +t')/2,\epsilon )$ around $S(t ,\epsilon )$ in the Wigner representation of Eq.\ \eqref{Wigner}, we obtain:
\begin{equation}
S(t ,t')=\frac{1}{2\pi \hbar }\int_{-\infty }^{\infty } \! \! \! \! d\epsilon
e^{-i\epsilon (t -t')/\hbar }e^{i \hbar \partial _{\epsilon }\partial
_{t }/2} S(t ,\epsilon ) \,  . 
\label{eq:Stwotimetransform}
\end{equation}
The current operator of Eq.\ \eqref{curop} can then be expressed as:
\begin{align}
I_\alpha(t) = & \frac{e}{2\pi\hbar} \sum_{n\beta m \gamma l} \int d\epsilon_1 \int d\epsilon_2 e^{i (\epsilon_1-\epsilon_2) t/\hbar} \times \nonumber \\
& \phi_{\alpha n \beta m \gamma l}(t,\epsilon_1,\epsilon_2) \hat{a}^\dag_{\beta m} (\epsilon_1) \hat{a}_{\gamma l}(\epsilon_2) \, ,
\label{curopPhi}
\end{align}
where
\begin{align}
& \phi_{\alpha n \beta m \gamma l}(t,\epsilon_1,\epsilon_2) = 
\delta_{\alpha n \beta m}  \delta_{\alpha n \gamma l}   \nonumber \\
&- e^{-i \hbar \partial _{\epsilon_1 }\partial
_{t }/2} S^*_{\alpha n \beta m} (t,\epsilon_1)
 e^{i \hbar \partial _{\epsilon_2 }\partial
_{t }/2} S_{\alpha n \gamma l}(t ,\epsilon_2 )
\, . 
 \end{align}

The current fluctuations are defined in Eq.\ \eqref{curfluct} and can be expressed as:
\begin{equation}
P_{\zeta \eta}(t_1,t_2) =  \frac{1}{2} \left[ F_{\zeta \eta} (t_1,t_2) + F_{\eta \zeta} (t_2,t_1) \right] \, 
\end{equation}
in terms of
\begin{equation}
F_{\zeta \eta} (t_1,t_2) = \langle I_\zeta(t_1) I_{\eta} (t_2) \rangle - \langle I_{\zeta}(t_1) \rangle \langle I_{\eta} (t_2) \rangle \, . 
\end{equation}
Using the expectation value of the fluctuations of Eq.\ \eqref{eq:fluct}, we find:
\begin{align}
& F_{\zeta \eta}  = \frac{e^2}{(2\pi \hbar)^2}  \sum_{n_\zeta n_\eta \beta m \gamma l}\int d\epsilon_1 \int d\epsilon_2   e^{i (\epsilon_1-\epsilon_2) (t_1-t_2)/\hbar} \times \nonumber  \\
& \phi_{\zeta n_{\zeta} \beta m \gamma l} (t_1,\epsilon_1,\epsilon_2) \phi_{\eta n_{\eta} \gamma l \beta m} (t_2,\epsilon_2,\epsilon_1)  f(\epsilon_1) [1-f(\epsilon_2)] \, . 
\label{Ftt}
\end{align}
We follow Ref.\ \onlinecite{Moskalets:PRB2004} (Eq.\ (9)), apart from a factor of 2, and define the zero-frequency noise as in Eq.\ \eqref{znoise}. We therefore introduce:
\begin{equation}
f_{\zeta \eta} = \int_0^T \frac{dt}{T}  \int_{-\infty}^\infty d\tau F_{\zeta \eta}(t + \tau/2,t - \tau/2) \,  
\label{fnoise}
\end{equation}
so that
\begin{equation}
p_{\zeta \eta} =  \left( f_{\zeta \eta} + f_{\eta \zeta} \right)/2 \, . 
\label{pfrel}
\end{equation}
We therefore first consider quantities of the form:
\begin{equation}
\lambda =  \int_0^T \frac{dt}{T} \int_{-\infty}^\infty d\tau e^{i (\epsilon_1-\epsilon_2)\tau/\hbar} A(t+\tau/2) B(t-\tau/2) \, ,
\label{lambda}
\end{equation}
where $A(t+\tau/2)$ and $B(t-\tau/2)$ are periodic functions with period $T$ that depend on the energies $\epsilon_1$ and $\epsilon_2$.
The Fourier transforms of the periodic functions are:
\begin{equation}
A(t) = \sum_n e^{i \omega_n t} A_n 
\end{equation}
and similarly for $B(t)$, where $\omega_n = n 2 \pi/T$ and $n$ is an integral number. The inverse transforms are defined in corresponding ways. We then obtain:
\begin{equation}
\lambda = 2 \pi \hbar \sum_n A_n B_{-n} \delta(\hbar \omega_n + (\epsilon_1-\epsilon_2)) \, . 
\end{equation}
Using Eqs.\ \eqref{Ftt} and \eqref{fnoise}, the low-frequency noise is of the form:
\begin{align}
\kappa &= \int d \epsilon_1 \int d\epsilon_2 f(\epsilon_1) \left[1 - f(\epsilon_2) \right] 2\pi \hbar \times \nonumber \\
& \sum_n A_n B_{-n} \delta(\hbar \omega_n + (\epsilon_1-\epsilon_2) \, ,
\end{align}
where $A_n$ and $B_{-n}$ depend on the energies $\epsilon_1$ and $\epsilon_2$. Carrying out the integral over the energy $\epsilon_2$:
\begin{equation}
\kappa = 2\pi \hbar \sum_n \int d\epsilon_1 f(\epsilon_1) \left[1 - f(\epsilon_1+\hbar \omega_n) \right] A_n B_{-n} \, , 
\end{equation}
where the energy $\epsilon_2=\epsilon_1+\hbar \omega_n$ in the coefficients $A_n$ and $B_{-n}$.

In metallic systems, the Fermi energy and exchange interaction are typically much larger than the driving frequency. In this case, we can approximate the scattering matrix as independent of the driving frequency $\hbar \omega_n$ and the temperature $k_B T$. We can then evaluate the scattering matrices at the Fermi energy and obtain:
\begin{align}
\kappa_n & = A_n(\epsilon_F) B_{-n}(\epsilon_F) 2 \pi \hbar \hbar \omega_n \left[1 + f_\text{BE}(\hbar \omega_n,k_B T) \right] \, , 
\end{align}
where the Bose-Einstein distribution function is:
\begin{equation}
f_\text{BE} = \frac{1}{\exp{\hbar \omega_n/k_B T}-1} \, .
\end{equation}
We then obtain:
\begin{equation}
f_{\zeta \eta} =   \sum_q Y_{\zeta \eta} (\omega_q) \hbar \omega_q \left[1 + f_{BE}(\hbar \omega_q) \right] \, , 
\end{equation}
where
\begin{equation}
Y_{\zeta \eta}(\hbar \omega_q) = \frac{e^2}{h} \sum_{n_\zeta n_\eta} \sum_{\beta m \gamma l} \phi_{\zeta n_\zeta \beta m \gamma l}(\omega_q) \phi_{\eta n_\eta \gamma l \beta m}(-\omega_q) \, ,
\end{equation}
and we have defined the Fourier transform as:
\begin{equation}
\phi_{\zeta n_\zeta \beta m \gamma l}(\omega_q) =\frac{1}{T} \int_0^T dt e^{-i\omega_q t} \phi_{\zeta n_\zeta \beta m \gamma l}(t,\epsilon_F,\epsilon_F) \, .
\end{equation}
We see that $Y_{\zeta \eta}(-\hbar \omega_q) = Y_{\eta \zeta}(\hbar \omega_q)$. Consequently, using Eq.\ \eqref{pfrel} and rewriting the Bose-Einstein distribution, we obtain:
\begin{align}
p_{\zeta \eta} & =\sum_q Y_{\zeta \eta}^{(s)}(\hbar \omega_q) k_B T \nonumber \\
& +\frac{1}{2} \sum_{q} Y_{\zeta \eta}^{(s)}(\hbar \omega_q) \left[ \hbar \omega_q \coth{\frac{\hbar \omega_q}{2 k_B T}} - 2k_B T \right] \, , 
\end{align}
where the first term represents the thermal noise contribution, the second represents the shot noise contribution, and
\begin{equation}
Y_{\zeta \eta}^{(s)}(\hbar \omega_q)=\left[ Y_{\zeta \eta}^{(s)}(\hbar \omega_q)+Y_{\zeta \eta}^{(s)}(-\hbar \omega_q) \right]/2 \, .
\end{equation}

At this point, consistent with our assumption that the scattering matrix is energy independent on the scale of the frequency and temperature, we can use the static scattering matrix in the evaluation of $Y$, $Y \rightarrow X$, where $X$ is defined in Eq.\ \eqref{X}. Hence, we obtain Eq.\ \eqref{crosscorr} with the quantities introduced in Eqs.\ \eqref{X}, \eqref{Phiq}, and \eqref{Phit}.

\bibliography{bibliography}

\begin{thebibliography}{66}%
\makeatletter
\providecommand \@ifxundefined [1]{%
 \@ifx{#1\undefined}
}%
\providecommand \@ifnum [1]{%
 \ifnum #1\expandafter \@firstoftwo
 \else \expandafter \@secondoftwo
 \fi
}%
\providecommand \@ifx [1]{%
 \ifx #1\expandafter \@firstoftwo
 \else \expandafter \@secondoftwo
 \fi
}%
\providecommand \natexlab [1]{#1}%
\providecommand \enquote  [1]{``#1''}%
\providecommand \bibnamefont  [1]{#1}%
\providecommand \bibfnamefont [1]{#1}%
\providecommand \citenamefont [1]{#1}%
\providecommand \href@noop [0]{\@secondoftwo}%
\providecommand \href [0]{\begingroup \@sanitize@url \@href}%
\providecommand \@href[1]{\@@startlink{#1}\@@href}%
\providecommand \@@href[1]{\endgroup#1\@@endlink}%
\providecommand \@sanitize@url [0]{\catcode `\\12\catcode `\$12\catcode
  `\&12\catcode `\#12\catcode `\^12\catcode `\_12\catcode `\%12\relax}%
\providecommand \@@startlink[1]{}%
\providecommand \@@endlink[0]{}%
\providecommand \url  [0]{\begingroup\@sanitize@url \@url }%
\providecommand \@url [1]{\endgroup\@href {#1}{\urlprefix }}%
\providecommand \urlprefix  [0]{URL }%
\providecommand \Eprint [0]{\href }%
\providecommand \doibase [0]{http://dx.doi.org/}%
\providecommand \selectlanguage [0]{\@gobble}%
\providecommand \bibinfo  [0]{\@secondoftwo}%
\providecommand \bibfield  [0]{\@secondoftwo}%
\providecommand \translation [1]{[#1]}%
\providecommand \BibitemOpen [0]{}%
\providecommand \bibitemStop [0]{}%
\providecommand \bibitemNoStop [0]{.\EOS\space}%
\providecommand \EOS [0]{\spacefactor3000\relax}%
\providecommand \BibitemShut  [1]{\csname bibitem#1\endcsname}%
\let\auto@bib@innerbib\@empty
\bibitem [{\citenamefont {Johnson}(1928)}]{Johnson:PR1928}%
  \BibitemOpen
  \bibfield  {author} {\bibinfo {author} {\bibfnamefont {J.~B.}\ \bibnamefont
  {Johnson}},\ }\href {\doibase 10.1103/PhysRev.32.97} {\bibfield  {journal}
  {\bibinfo  {journal} {Physical Review}\ }\textbf {\bibinfo {volume} {32}},\
  \bibinfo {pages} {97} (\bibinfo {year} {1928})}\BibitemShut {NoStop}%
\bibitem [{\citenamefont {Nyquist}(1928)}]{Nyquist:PR1928}%
  \BibitemOpen
  \bibfield  {author} {\bibinfo {author} {\bibfnamefont {H.}~\bibnamefont
  {Nyquist}},\ }\href {http://link.aps.org/abstract/PR/v32/p110} {\bibfield
  {journal} {\bibinfo  {journal} {Physical Review}\ }\textbf {\bibinfo {volume}
  {32}},\ \bibinfo {pages} {110} (\bibinfo {year} {1928})}\BibitemShut
  {NoStop}%
\bibitem [{\citenamefont {Blanter}\ and\ \citenamefont
  {B\"{u}ttiker}(2000)}]{Blanter:PhysRep2000}%
  \BibitemOpen
  \bibfield  {author} {\bibinfo {author} {\bibfnamefont {Y.~M.}\ \bibnamefont
  {Blanter}}\ and\ \bibinfo {author} {\bibfnamefont {M.}~\bibnamefont
  {B\"{u}ttiker}},\ }\href {\doibase
  http://dx.doi.org/10.1016/S0370-1573(99)00123-4} {\bibfield  {journal}
  {\bibinfo  {journal} {Physics Reports}\ }\textbf {\bibinfo {volume} {336}},\
  \bibinfo {pages} {1} (\bibinfo {year} {2000})}\BibitemShut {NoStop}%
\bibitem [{\citenamefont {Berger}(1996)}]{Berger:PRB1996}%
  \BibitemOpen
  \bibfield  {author} {\bibinfo {author} {\bibfnamefont {L.}~\bibnamefont
  {Berger}},\ }\href {http://link.aps.org/abstract/PRB/v54/p9353} {\bibfield
  {journal} {\bibinfo  {journal} {Physical Review B}\ }\textbf {\bibinfo
  {volume} {54}},\ \bibinfo {pages} {9353} (\bibinfo {year}
  {1996})}\BibitemShut {NoStop}%
\bibitem [{\citenamefont {Slonczewski}(1996)}]{Slonczewski:JMMM1996}%
  \BibitemOpen
  \bibfield  {author} {\bibinfo {author} {\bibfnamefont {J.~C.}\ \bibnamefont
  {Slonczewski}},\ }\href {<Go to ISI>://A1996VH55200001} {\bibfield  {journal}
  {\bibinfo  {journal} {Journal of Magnetism and Magnetic Materials}\ }\textbf
  {\bibinfo {volume} {159}},\ \bibinfo {pages} {L1} (\bibinfo {year}
  {1996})}\BibitemShut {NoStop}%
\bibitem [{\citenamefont {Tsoi}\ \emph {et~al.}(1998)\citenamefont {Tsoi},
  \citenamefont {Jansen}, \citenamefont {Bass}, \citenamefont {Chiang},
  \citenamefont {Seck}, \citenamefont {Tsoi},\ and\ \citenamefont
  {Wyder}}]{Tsoi:PRL1998}%
  \BibitemOpen
  \bibfield  {author} {\bibinfo {author} {\bibfnamefont {M.}~\bibnamefont
  {Tsoi}}, \bibinfo {author} {\bibfnamefont {A.~G.~M.}\ \bibnamefont {Jansen}},
  \bibinfo {author} {\bibfnamefont {J.}~\bibnamefont {Bass}}, \bibinfo {author}
  {\bibfnamefont {W.~C.}\ \bibnamefont {Chiang}}, \bibinfo {author}
  {\bibfnamefont {M.}~\bibnamefont {Seck}}, \bibinfo {author} {\bibfnamefont
  {V.}~\bibnamefont {Tsoi}}, \ and\ \bibinfo {author} {\bibfnamefont
  {P.}~\bibnamefont {Wyder}},\ }\href
  {http://link.aps.org/doi/10.1103/PhysRevLett.80.4281} {\bibfield  {journal}
  {\bibinfo  {journal} {Physical Review Letters}\ }\textbf {\bibinfo {volume}
  {80}},\ \bibinfo {pages} {4281} (\bibinfo {year} {1998})}\BibitemShut
  {NoStop}%
\bibitem [{\citenamefont {Myers}\ \emph {et~al.}(1999)\citenamefont {Myers},
  \citenamefont {Ralph}, \citenamefont {Katine}, \citenamefont {Louie},\ and\
  \citenamefont {Buhrman}}]{Myers:Science1999}%
  \BibitemOpen
  \bibfield  {author} {\bibinfo {author} {\bibfnamefont {E.~B.}\ \bibnamefont
  {Myers}}, \bibinfo {author} {\bibfnamefont {D.~C.}\ \bibnamefont {Ralph}},
  \bibinfo {author} {\bibfnamefont {J.~A.}\ \bibnamefont {Katine}}, \bibinfo
  {author} {\bibfnamefont {R.~N.}\ \bibnamefont {Louie}}, \ and\ \bibinfo
  {author} {\bibfnamefont {R.~A.}\ \bibnamefont {Buhrman}},\ }\href {\doibase
  10.1126/science.285.5429.867} {\bibfield  {journal} {\bibinfo  {journal}
  {Science}\ }\textbf {\bibinfo {volume} {285}},\ \bibinfo {pages} {867}
  (\bibinfo {year} {1999})}\BibitemShut {NoStop}%
\bibitem [{\citenamefont {Katine}\ \emph {et~al.}(2000)\citenamefont {Katine},
  \citenamefont {Albert}, \citenamefont {Buhrman}, \citenamefont {Myers},\ and\
  \citenamefont {Ralph}}]{Katine:PRL2000}%
  \BibitemOpen
  \bibfield  {author} {\bibinfo {author} {\bibfnamefont {J.~A.}\ \bibnamefont
  {Katine}}, \bibinfo {author} {\bibfnamefont {F.~J.}\ \bibnamefont {Albert}},
  \bibinfo {author} {\bibfnamefont {R.~A.}\ \bibnamefont {Buhrman}}, \bibinfo
  {author} {\bibfnamefont {E.~B.}\ \bibnamefont {Myers}}, \ and\ \bibinfo
  {author} {\bibfnamefont {D.~C.}\ \bibnamefont {Ralph}},\ }\href {<Go to
  ISI>://000086243800032} {\bibfield  {journal} {\bibinfo  {journal} {Physical
  Review Letters}\ }\textbf {\bibinfo {volume} {84}},\ \bibinfo {pages} {3149}
  (\bibinfo {year} {2000})}\BibitemShut {NoStop}%
\bibitem [{\citenamefont {Kiselev}\ \emph {et~al.}(2003)\citenamefont
  {Kiselev}, \citenamefont {Sankey}, \citenamefont {Krivorotov}, \citenamefont
  {Emley}, \citenamefont {Schoelkopf}, \citenamefont {Buhrman},\ and\
  \citenamefont {Ralph}}]{Kiselev:Nature2003}%
  \BibitemOpen
  \bibfield  {author} {\bibinfo {author} {\bibfnamefont {S.~I.}\ \bibnamefont
  {Kiselev}}, \bibinfo {author} {\bibfnamefont {J.~C.}\ \bibnamefont {Sankey}},
  \bibinfo {author} {\bibfnamefont {I.~N.}\ \bibnamefont {Krivorotov}},
  \bibinfo {author} {\bibfnamefont {N.~C.}\ \bibnamefont {Emley}}, \bibinfo
  {author} {\bibfnamefont {R.~J.}\ \bibnamefont {Schoelkopf}}, \bibinfo
  {author} {\bibfnamefont {R.~A.}\ \bibnamefont {Buhrman}}, \ and\ \bibinfo
  {author} {\bibfnamefont {D.~C.}\ \bibnamefont {Ralph}},\ }\href
  {http://dx.doi.org/10.1038/nature01967} {\bibfield  {journal} {\bibinfo
  {journal} {Nature}\ }\textbf {\bibinfo {volume} {425}},\ \bibinfo {pages}
  {380} (\bibinfo {year} {2003})}\BibitemShut {NoStop}%
\bibitem [{\citenamefont {Krivorotov}\ \emph {et~al.}(2005)\citenamefont
  {Krivorotov}, \citenamefont {Emley}, \citenamefont {Sankey}, \citenamefont
  {Kiselev}, \citenamefont {Ralph},\ and\ \citenamefont
  {Buhrman}}]{Krivorotov:Science2005}%
  \BibitemOpen
  \bibfield  {author} {\bibinfo {author} {\bibfnamefont {I.~N.}\ \bibnamefont
  {Krivorotov}}, \bibinfo {author} {\bibfnamefont {N.~C.}\ \bibnamefont
  {Emley}}, \bibinfo {author} {\bibfnamefont {J.~C.}\ \bibnamefont {Sankey}},
  \bibinfo {author} {\bibfnamefont {S.~I.}\ \bibnamefont {Kiselev}}, \bibinfo
  {author} {\bibfnamefont {D.~C.}\ \bibnamefont {Ralph}}, \ and\ \bibinfo
  {author} {\bibfnamefont {R.~A.}\ \bibnamefont {Buhrman}},\ }\href {<Go to
  ISI>://000226361900034} {\bibfield  {journal} {\bibinfo  {journal} {Science}\
  }\textbf {\bibinfo {volume} {307}},\ \bibinfo {pages} {228} (\bibinfo {year}
  {2005})}\BibitemShut {NoStop}%
\bibitem [{\citenamefont {Manchon}\ and\ \citenamefont
  {Zhang}(2008)}]{Manchon:PRB2008}%
  \BibitemOpen
  \bibfield  {author} {\bibinfo {author} {\bibfnamefont {A.}~\bibnamefont
  {Manchon}}\ and\ \bibinfo {author} {\bibfnamefont {S.}~\bibnamefont
  {Zhang}},\ }\href {http://link.aps.org/abstract/PRB/v78/e212405} {\bibfield
  {journal} {\bibinfo  {journal} {Physical Review B (Condensed Matter and
  Materials Physics)}\ }\textbf {\bibinfo {volume} {78}},\ \bibinfo {pages}
  {212405} (\bibinfo {year} {2008})}\BibitemShut {NoStop}%
\bibitem [{\citenamefont {Miron}\ \emph {et~al.}(2010)\citenamefont {Miron},
  \citenamefont {Gaudin}, \citenamefont {Auffret}, \citenamefont {Rodmacq},
  \citenamefont {Schuhl}, \citenamefont {Pizzini}, \citenamefont {Vogel},\ and\
  \citenamefont {Gambardella}}]{Miron:NatMat2010}%
  \BibitemOpen
  \bibfield  {author} {\bibinfo {author} {\bibfnamefont {I.~M.}\ \bibnamefont
  {Miron}}, \bibinfo {author} {\bibfnamefont {G.}~\bibnamefont {Gaudin}},
  \bibinfo {author} {\bibfnamefont {S.}~\bibnamefont {Auffret}}, \bibinfo
  {author} {\bibfnamefont {B.}~\bibnamefont {Rodmacq}}, \bibinfo {author}
  {\bibfnamefont {A.}~\bibnamefont {Schuhl}}, \bibinfo {author} {\bibfnamefont
  {S.}~\bibnamefont {Pizzini}}, \bibinfo {author} {\bibfnamefont
  {J.}~\bibnamefont {Vogel}}, \ and\ \bibinfo {author} {\bibfnamefont
  {P.}~\bibnamefont {Gambardella}},\ }\href
  {http://dx.doi.org/10.1038/nmat2613} {\bibfield  {journal} {\bibinfo
  {journal} {Nature Materials}\ }\textbf {\bibinfo {volume} {9}},\ \bibinfo
  {pages} {230} (\bibinfo {year} {2010})}\BibitemShut {NoStop}%
\bibitem [{\citenamefont {Miron}\ \emph {et~al.}(2011)\citenamefont {Miron},
  \citenamefont {Garello}, \citenamefont {Gaudin}, \citenamefont {Zermatten},
  \citenamefont {Costache}, \citenamefont {Auffret}, \citenamefont {Bandiera},
  \citenamefont {Rodmacq}, \citenamefont {Schuhl},\ and\ \citenamefont
  {Gambardella}}]{Miron:Nature2011}%
  \BibitemOpen
  \bibfield  {author} {\bibinfo {author} {\bibfnamefont {I.~M.}\ \bibnamefont
  {Miron}}, \bibinfo {author} {\bibfnamefont {K.}~\bibnamefont {Garello}},
  \bibinfo {author} {\bibfnamefont {G.}~\bibnamefont {Gaudin}}, \bibinfo
  {author} {\bibfnamefont {P.-J.}\ \bibnamefont {Zermatten}}, \bibinfo {author}
  {\bibfnamefont {M.~V.}\ \bibnamefont {Costache}}, \bibinfo {author}
  {\bibfnamefont {S.}~\bibnamefont {Auffret}}, \bibinfo {author} {\bibfnamefont
  {S.}~\bibnamefont {Bandiera}}, \bibinfo {author} {\bibfnamefont
  {B.}~\bibnamefont {Rodmacq}}, \bibinfo {author} {\bibfnamefont
  {A.}~\bibnamefont {Schuhl}}, \ and\ \bibinfo {author} {\bibfnamefont
  {P.}~\bibnamefont {Gambardella}},\ }\href {\doibase
  http://www.nature.com/nature/journal/v476/n7359/abs/nature10309.html#supplementary-information}
  {\bibfield  {journal} {\bibinfo  {journal} {Nature}\ }\textbf {\bibinfo
  {volume} {476}},\ \bibinfo {pages} {189} (\bibinfo {year}
  {2011})}\BibitemShut {NoStop}%
\bibitem [{\citenamefont {Liu}\ \emph {et~al.}(2012)\citenamefont {Liu},
  \citenamefont {Pai}, \citenamefont {Li}, \citenamefont {Tseng}, \citenamefont
  {Ralph},\ and\ \citenamefont {Buhrman}}]{Liu:Science2012}%
  \BibitemOpen
  \bibfield  {author} {\bibinfo {author} {\bibfnamefont {L.}~\bibnamefont
  {Liu}}, \bibinfo {author} {\bibfnamefont {C.-F.}\ \bibnamefont {Pai}},
  \bibinfo {author} {\bibfnamefont {Y.}~\bibnamefont {Li}}, \bibinfo {author}
  {\bibfnamefont {H.~W.}\ \bibnamefont {Tseng}}, \bibinfo {author}
  {\bibfnamefont {D.~C.}\ \bibnamefont {Ralph}}, \ and\ \bibinfo {author}
  {\bibfnamefont {R.~A.}\ \bibnamefont {Buhrman}},\ }\href {\doibase
  10.1126/science.1218197} {\bibfield  {journal} {\bibinfo  {journal}
  {Science}\ }\textbf {\bibinfo {volume} {336}},\ \bibinfo {pages} {555}
  (\bibinfo {year} {2012})}\BibitemShut {NoStop}%
\bibitem [{\citenamefont {Garello}\ \emph {et~al.}(2013)\citenamefont
  {Garello}, \citenamefont {Miron}, \citenamefont {Avci}, \citenamefont
  {Freimuth}, \citenamefont {Mokrousov}, \citenamefont {Blugel}, \citenamefont
  {Auffret}, \citenamefont {Boulle}, \citenamefont {Gaudin},\ and\
  \citenamefont {Gambardella}}]{Garello:NatNan2013}%
  \BibitemOpen
  \bibfield  {author} {\bibinfo {author} {\bibfnamefont {K.}~\bibnamefont
  {Garello}}, \bibinfo {author} {\bibfnamefont {I.~M.}\ \bibnamefont {Miron}},
  \bibinfo {author} {\bibfnamefont {C.~O.}\ \bibnamefont {Avci}}, \bibinfo
  {author} {\bibfnamefont {F.}~\bibnamefont {Freimuth}}, \bibinfo {author}
  {\bibfnamefont {Y.}~\bibnamefont {Mokrousov}}, \bibinfo {author}
  {\bibfnamefont {S.}~\bibnamefont {Blugel}}, \bibinfo {author} {\bibfnamefont
  {S.}~\bibnamefont {Auffret}}, \bibinfo {author} {\bibfnamefont
  {O.}~\bibnamefont {Boulle}}, \bibinfo {author} {\bibfnamefont
  {G.}~\bibnamefont {Gaudin}}, \ and\ \bibinfo {author} {\bibfnamefont
  {P.}~\bibnamefont {Gambardella}},\ }\href
  {http://dx.doi.org/10.1038/nnano.2013.145} {\bibfield  {journal} {\bibinfo
  {journal} {Nature Nanotechnology}\ }\textbf {\bibinfo {volume} {8}},\
  \bibinfo {pages} {587} (\bibinfo {year} {2013})}\BibitemShut {NoStop}%
\bibitem [{\citenamefont {Mellnik}\ \emph {et~al.}(2014)\citenamefont
  {Mellnik}, \citenamefont {Lee}, \citenamefont {Richardella}, \citenamefont
  {Grab}, \citenamefont {Mintun}, \citenamefont {Fischer}, \citenamefont
  {Vaezi}, \citenamefont {Manchon}, \citenamefont {Kim}, \citenamefont
  {Samarth},\ and\ \citenamefont {Ralph}}]{Mellnik:Nature2014}%
  \BibitemOpen
  \bibfield  {author} {\bibinfo {author} {\bibfnamefont {A.~R.}\ \bibnamefont
  {Mellnik}}, \bibinfo {author} {\bibfnamefont {J.~S.}\ \bibnamefont {Lee}},
  \bibinfo {author} {\bibfnamefont {A.}~\bibnamefont {Richardella}}, \bibinfo
  {author} {\bibfnamefont {J.~L.}\ \bibnamefont {Grab}}, \bibinfo {author}
  {\bibfnamefont {P.~J.}\ \bibnamefont {Mintun}}, \bibinfo {author}
  {\bibfnamefont {M.~H.}\ \bibnamefont {Fischer}}, \bibinfo {author}
  {\bibfnamefont {A.}~\bibnamefont {Vaezi}}, \bibinfo {author} {\bibfnamefont
  {A.}~\bibnamefont {Manchon}}, \bibinfo {author} {\bibfnamefont {E.~A.}\
  \bibnamefont {Kim}}, \bibinfo {author} {\bibfnamefont {N.}~\bibnamefont
  {Samarth}}, \ and\ \bibinfo {author} {\bibfnamefont {D.~C.}\ \bibnamefont
  {Ralph}},\ }\href {\doibase 10.1038/nature13534} {\bibfield  {journal}
  {\bibinfo  {journal} {Nature}\ }\textbf {\bibinfo {volume} {511}},\ \bibinfo
  {pages} {449} (\bibinfo {year} {2014})}\BibitemShut {NoStop}%
\bibitem [{\citenamefont {MacNeill}\ \emph {et~al.}(2017)\citenamefont
  {MacNeill}, \citenamefont {Stiehl}, \citenamefont {Guimaraes}, \citenamefont
  {Buhrman}, \citenamefont {Park},\ and\ \citenamefont
  {Ralph}}]{MacNeill:NatPhys2017}%
  \BibitemOpen
  \bibfield  {author} {\bibinfo {author} {\bibfnamefont {D.}~\bibnamefont
  {MacNeill}}, \bibinfo {author} {\bibfnamefont {G.~M.}\ \bibnamefont
  {Stiehl}}, \bibinfo {author} {\bibfnamefont {M.~H.~D.}\ \bibnamefont
  {Guimaraes}}, \bibinfo {author} {\bibfnamefont {R.~A.}\ \bibnamefont
  {Buhrman}}, \bibinfo {author} {\bibfnamefont {J.}~\bibnamefont {Park}}, \
  and\ \bibinfo {author} {\bibfnamefont {D.~C.}\ \bibnamefont {Ralph}},\ }\href
  {\doibase 10.1038/nphys3933
  http://www.nature.com/nphys/journal/v13/n3/abs/nphys3933.html#supplementary-information}
  {\bibfield  {journal} {\bibinfo  {journal} {Nature Physics}\ }\textbf
  {\bibinfo {volume} {13}},\ \bibinfo {pages} {300} (\bibinfo {year}
  {2017})}\BibitemShut {NoStop}%
\bibitem [{\citenamefont {Zhu}\ \emph {et~al.}(2019)\citenamefont {Zhu},
  \citenamefont {Ralph},\ and\ \citenamefont {Buhrman}}]{Zhu:PRL2019}%
  \BibitemOpen
  \bibfield  {author} {\bibinfo {author} {\bibfnamefont {L.}~\bibnamefont
  {Zhu}}, \bibinfo {author} {\bibfnamefont {D.~C.}\ \bibnamefont {Ralph}}, \
  and\ \bibinfo {author} {\bibfnamefont {R.~A.}\ \bibnamefont {Buhrman}},\
  }\href {\doibase 10.1103/PhysRevLett.122.077201} {\bibfield  {journal}
  {\bibinfo  {journal} {Physical Review Letters}\ }\textbf {\bibinfo {volume}
  {122}},\ \bibinfo {pages} {077201} (\bibinfo {year} {2019})}\BibitemShut
  {NoStop}%
\bibitem [{\citenamefont {Silva}\ and\ \citenamefont
  {Rippard}(2008)}]{Silva:JMMM2008}%
  \BibitemOpen
  \bibfield  {author} {\bibinfo {author} {\bibfnamefont {T.~J.}\ \bibnamefont
  {Silva}}\ and\ \bibinfo {author} {\bibfnamefont {W.~H.}\ \bibnamefont
  {Rippard}},\ }\href {\doibase 10.1016/j.jmmm.2007.12.022} {\bibfield
  {journal} {\bibinfo  {journal} {Journal of Magnetism and Magnetic Materials}\
  }\textbf {\bibinfo {volume} {320}},\ \bibinfo {pages} {1260} (\bibinfo {year}
  {2008})}\BibitemShut {NoStop}%
\bibitem [{\citenamefont {Ralph}\ and\ \citenamefont
  {Stiles}(2008)}]{Ralph:JMMM2008}%
  \BibitemOpen
  \bibfield  {author} {\bibinfo {author} {\bibfnamefont {D.~C.}\ \bibnamefont
  {Ralph}}\ and\ \bibinfo {author} {\bibfnamefont {M.~D.}\ \bibnamefont
  {Stiles}},\ }\href
  {http://www.sciencedirect.com/science/article/B6TJJ-4RFSD1M-2/2/f35a2bc5e9c53f19f6883d74c20dbb69}
  {\bibfield  {journal} {\bibinfo  {journal} {Journal of Magnetism and Magnetic
  Materials}\ }\textbf {\bibinfo {volume} {320}},\ \bibinfo {pages} {1190}
  (\bibinfo {year} {2008})}\BibitemShut {NoStop}%
\bibitem [{\citenamefont {Brataas}\ \emph {et~al.}(2012)\citenamefont
  {Brataas}, \citenamefont {Kent},\ and\ \citenamefont
  {Ohno}}]{Brataas:NatMat2012}%
  \BibitemOpen
  \bibfield  {author} {\bibinfo {author} {\bibfnamefont {A.}~\bibnamefont
  {Brataas}}, \bibinfo {author} {\bibfnamefont {A.~D.}\ \bibnamefont {Kent}}, \
  and\ \bibinfo {author} {\bibfnamefont {H.}~\bibnamefont {Ohno}},\ }\href
  {\doibase 10.1038/nmat3311} {\bibfield  {journal} {\bibinfo  {journal}
  {Nature Materials}\ }\textbf {\bibinfo {volume} {11}},\ \bibinfo {pages}
  {372} (\bibinfo {year} {2012})}\BibitemShut {NoStop}%
\bibitem [{\citenamefont {Manchon}\ \emph {et~al.}(2019)\citenamefont
  {Manchon}, \citenamefont {\v{Z}elezn\`{y}}, \citenamefont {Miron},
  \citenamefont {Jungwirth}, \citenamefont {Sinova}, \citenamefont {Thiaville},
  \citenamefont {Garello},\ and\ \citenamefont
  {Gambardella}}]{Manchon:RMP2019}%
  \BibitemOpen
  \bibfield  {author} {\bibinfo {author} {\bibfnamefont {A.}~\bibnamefont
  {Manchon}}, \bibinfo {author} {\bibfnamefont {J.}~\bibnamefont
  {\v{Z}elezn\`{y}}}, \bibinfo {author} {\bibfnamefont {I.~M.}\ \bibnamefont
  {Miron}}, \bibinfo {author} {\bibfnamefont {T.}~\bibnamefont {Jungwirth}},
  \bibinfo {author} {\bibfnamefont {J.}~\bibnamefont {Sinova}}, \bibinfo
  {author} {\bibfnamefont {A.}~\bibnamefont {Thiaville}}, \bibinfo {author}
  {\bibfnamefont {K.}~\bibnamefont {Garello}}, \ and\ \bibinfo {author}
  {\bibfnamefont {P.}~\bibnamefont {Gambardella}},\ }\href {\doibase
  10.1103/RevModPhys.91.035004} {\bibfield  {journal} {\bibinfo  {journal}
  {Reviews of Modern Physics}\ }\textbf {\bibinfo {volume} {91}},\ \bibinfo
  {pages} {035004} (\bibinfo {year} {2019})}\BibitemShut {NoStop}%
\bibitem [{\citenamefont {Monod}\ \emph {et~al.}(1972)\citenamefont {Monod},
  \citenamefont {Hurdequint}, \citenamefont {Janossy}, \citenamefont {Obert},\
  and\ \citenamefont {Chaumont}}]{Monod:PRL1972}%
  \BibitemOpen
  \bibfield  {author} {\bibinfo {author} {\bibfnamefont {P.}~\bibnamefont
  {Monod}}, \bibinfo {author} {\bibfnamefont {H.}~\bibnamefont {Hurdequint}},
  \bibinfo {author} {\bibfnamefont {A.}~\bibnamefont {Janossy}}, \bibinfo
  {author} {\bibfnamefont {J.}~\bibnamefont {Obert}}, \ and\ \bibinfo {author}
  {\bibfnamefont {J.}~\bibnamefont {Chaumont}},\ }\href
  {http://link.aps.org/doi/10.1103/PhysRevLett.29.1327} {\bibfield  {journal}
  {\bibinfo  {journal} {Physical Review Letters}\ }\textbf {\bibinfo {volume}
  {29}},\ \bibinfo {pages} {1327} (\bibinfo {year} {1972})}\BibitemShut
  {NoStop}%
\bibitem [{\citenamefont {Silsbee}\ \emph {et~al.}(1979)\citenamefont
  {Silsbee}, \citenamefont {Janossy},\ and\ \citenamefont
  {Monod}}]{Silsbee:PRB1979}%
  \BibitemOpen
  \bibfield  {author} {\bibinfo {author} {\bibfnamefont {R.~H.}\ \bibnamefont
  {Silsbee}}, \bibinfo {author} {\bibfnamefont {A.}~\bibnamefont {Janossy}}, \
  and\ \bibinfo {author} {\bibfnamefont {P.}~\bibnamefont {Monod}},\ }\href
  {http://link.aps.org/doi/10.1103/PhysRevB.19.4382} {\bibfield  {journal}
  {\bibinfo  {journal} {Physical Review B}\ }\textbf {\bibinfo {volume} {19}},\
  \bibinfo {pages} {4382} (\bibinfo {year} {1979})}\BibitemShut {NoStop}%
\bibitem [{\citenamefont {Urban}\ \emph {et~al.}(2001)\citenamefont {Urban},
  \citenamefont {Woltersdorf},\ and\ \citenamefont {Heinrich}}]{Urban:PRL2001}%
  \BibitemOpen
  \bibfield  {author} {\bibinfo {author} {\bibfnamefont {R.}~\bibnamefont
  {Urban}}, \bibinfo {author} {\bibfnamefont {G.}~\bibnamefont {Woltersdorf}},
  \ and\ \bibinfo {author} {\bibfnamefont {B.}~\bibnamefont {Heinrich}},\
  }\href {http://link.aps.org/doi/10.1103/PhysRevLett.87.217204} {\bibfield
  {journal} {\bibinfo  {journal} {Physical Review Letters}\ }\textbf {\bibinfo
  {volume} {87}},\ \bibinfo {pages} {217204} (\bibinfo {year}
  {2001})}\BibitemShut {NoStop}%
\bibitem [{\citenamefont {Mizukami}\ \emph {et~al.}(2001)\citenamefont
  {Mizukami}, \citenamefont {Ando},\ and\ \citenamefont
  {Miazaki}}]{Mizukami:Jpn2001}%
  \BibitemOpen
  \bibfield  {author} {\bibinfo {author} {\bibfnamefont {S.}~\bibnamefont
  {Mizukami}}, \bibinfo {author} {\bibfnamefont {Y.}~\bibnamefont {Ando}}, \
  and\ \bibinfo {author} {\bibfnamefont {T.}~\bibnamefont {Miazaki}},\
  }\href@noop {} {\bibfield  {journal} {\bibinfo  {journal} {Jpn. J. Appl.
  Phys.}\ }\textbf {\bibinfo {volume} {40}},\ \bibinfo {pages} {580} (\bibinfo
  {year} {2001})}\BibitemShut {NoStop}%
\bibitem [{\citenamefont {Tserkovnyak}\ \emph {et~al.}(2002)\citenamefont
  {Tserkovnyak}, \citenamefont {Brataas},\ and\ \citenamefont
  {Bauer}}]{Tserkovnyak:PRL2001}%
  \BibitemOpen
  \bibfield  {author} {\bibinfo {author} {\bibfnamefont {Y.}~\bibnamefont
  {Tserkovnyak}}, \bibinfo {author} {\bibfnamefont {A.}~\bibnamefont
  {Brataas}}, \ and\ \bibinfo {author} {\bibfnamefont {G.~E.~W.}\ \bibnamefont
  {Bauer}},\ }\href@noop {} {\bibfield  {journal} {\bibinfo  {journal}
  {Physical Review Letters}\ }\textbf {\bibinfo {volume} {88}},\ \bibinfo
  {pages} {117601} (\bibinfo {year} {2002})}\BibitemShut {NoStop}%
\bibitem [{\citenamefont {Heinrich}\ \emph {et~al.}(2003)\citenamefont
  {Heinrich}, \citenamefont {Tserkovnyak}, \citenamefont {Woltersdorf},
  \citenamefont {Brataas}, \citenamefont {Urban},\ and\ \citenamefont
  {Bauer}}]{Heinrich:PRL2003}%
  \BibitemOpen
  \bibfield  {author} {\bibinfo {author} {\bibfnamefont {B.}~\bibnamefont
  {Heinrich}}, \bibinfo {author} {\bibfnamefont {Y.}~\bibnamefont
  {Tserkovnyak}}, \bibinfo {author} {\bibfnamefont {G.}~\bibnamefont
  {Woltersdorf}}, \bibinfo {author} {\bibfnamefont {A.}~\bibnamefont
  {Brataas}}, \bibinfo {author} {\bibfnamefont {R.}~\bibnamefont {Urban}}, \
  and\ \bibinfo {author} {\bibfnamefont {G.~E.~W.}\ \bibnamefont {Bauer}},\
  }\href@noop {} {\bibfield  {journal} {\bibinfo  {journal} {Physical Review
  Letters}\ }\textbf {\bibinfo {volume} {90}},\ \bibinfo {pages} {187601}
  (\bibinfo {year} {2003})}\BibitemShut {NoStop}%
\bibitem [{\citenamefont {Mosendz}\ \emph {et~al.}(2010)\citenamefont
  {Mosendz}, \citenamefont {Vlaminck}, \citenamefont {Pearson}, \citenamefont
  {Fradin}, \citenamefont {Bauer}, \citenamefont {Bader},\ and\ \citenamefont
  {Hoffmann}}]{Mosendz:PRB2010}%
  \BibitemOpen
  \bibfield  {author} {\bibinfo {author} {\bibfnamefont {O.}~\bibnamefont
  {Mosendz}}, \bibinfo {author} {\bibfnamefont {V.}~\bibnamefont {Vlaminck}},
  \bibinfo {author} {\bibfnamefont {J.~E.}\ \bibnamefont {Pearson}}, \bibinfo
  {author} {\bibfnamefont {F.~Y.}\ \bibnamefont {Fradin}}, \bibinfo {author}
  {\bibfnamefont {G.~E.~W.}\ \bibnamefont {Bauer}}, \bibinfo {author}
  {\bibfnamefont {S.~D.}\ \bibnamefont {Bader}}, \ and\ \bibinfo {author}
  {\bibfnamefont {A.}~\bibnamefont {Hoffmann}},\ }\href
  {http://link.aps.org/doi/10.1103/PhysRevB.82.214403} {\bibfield  {journal}
  {\bibinfo  {journal} {Physical Review B}\ }\textbf {\bibinfo {volume} {82}},\
  \bibinfo {pages} {214403} (\bibinfo {year} {2010})}\BibitemShut {NoStop}%
\bibitem [{\citenamefont {Jungwirth}\ \emph {et~al.}(2016)\citenamefont
  {Jungwirth}, \citenamefont {Marti}, \citenamefont {Wadley},\ and\
  \citenamefont {Wunderlich}}]{Jungwirth:NatNano2016}%
  \BibitemOpen
  \bibfield  {author} {\bibinfo {author} {\bibfnamefont {T.}~\bibnamefont
  {Jungwirth}}, \bibinfo {author} {\bibfnamefont {X.}~\bibnamefont {Marti}},
  \bibinfo {author} {\bibfnamefont {P.}~\bibnamefont {Wadley}}, \ and\ \bibinfo
  {author} {\bibfnamefont {J.}~\bibnamefont {Wunderlich}},\ }\href {\doibase
  10.1038/nnano.2016.18} {\bibfield  {journal} {\bibinfo  {journal} {Nat Nano}\
  }\textbf {\bibinfo {volume} {11}},\ \bibinfo {pages} {231} (\bibinfo {year}
  {2016})}\BibitemShut {NoStop}%
\bibitem [{\citenamefont {Bodnar}\ \emph {et~al.}(2018)\citenamefont {Bodnar},
  \citenamefont {Smejkal}, \citenamefont {Turek}, \citenamefont {Jungwirth},
  \citenamefont {Gomonay}, \citenamefont {Sinova}, \citenamefont {Sapozhnik},
  \citenamefont {Elmers}, \citenamefont {Kl\"{a}ui},\ and\ \citenamefont
  {Jourdan}}]{Bodnar:NatCom2018}%
  \BibitemOpen
  \bibfield  {author} {\bibinfo {author} {\bibfnamefont {S.~Y.}\ \bibnamefont
  {Bodnar}}, \bibinfo {author} {\bibfnamefont {L.}~\bibnamefont {Smejkal}},
  \bibinfo {author} {\bibfnamefont {I.}~\bibnamefont {Turek}}, \bibinfo
  {author} {\bibfnamefont {T.}~\bibnamefont {Jungwirth}}, \bibinfo {author}
  {\bibfnamefont {O.}~\bibnamefont {Gomonay}}, \bibinfo {author} {\bibfnamefont
  {J.}~\bibnamefont {Sinova}}, \bibinfo {author} {\bibfnamefont {A.~A.}\
  \bibnamefont {Sapozhnik}}, \bibinfo {author} {\bibfnamefont {H.~J.}\
  \bibnamefont {Elmers}}, \bibinfo {author} {\bibfnamefont {M.}~\bibnamefont
  {Kl\"{a}ui}}, \ and\ \bibinfo {author} {\bibfnamefont {M.}~\bibnamefont
  {Jourdan}},\ }\href {\doibase 10.1038/s41467-017-02780-x} {\bibfield
  {journal} {\bibinfo  {journal} {Nature Communications}\ }\textbf {\bibinfo
  {volume} {9}},\ \bibinfo {pages} {348} (\bibinfo {year} {2018})}\BibitemShut
  {NoStop}%
\bibitem [{\citenamefont {Lebrun}\ \emph {et~al.}(2018)\citenamefont {Lebrun},
  \citenamefont {Ross}, \citenamefont {Bender}, \citenamefont {Qaiumzadeh},
  \citenamefont {Baldrati}, \citenamefont {Cramer}, \citenamefont {Brataas},
  \citenamefont {Duine},\ and\ \citenamefont {Kl\"{a}ui}}]{Lebrun:Nature2018}%
  \BibitemOpen
  \bibfield  {author} {\bibinfo {author} {\bibfnamefont {R.}~\bibnamefont
  {Lebrun}}, \bibinfo {author} {\bibfnamefont {A.}~\bibnamefont {Ross}},
  \bibinfo {author} {\bibfnamefont {S.~A.}\ \bibnamefont {Bender}}, \bibinfo
  {author} {\bibfnamefont {A.}~\bibnamefont {Qaiumzadeh}}, \bibinfo {author}
  {\bibfnamefont {L.}~\bibnamefont {Baldrati}}, \bibinfo {author}
  {\bibfnamefont {J.}~\bibnamefont {Cramer}}, \bibinfo {author} {\bibfnamefont
  {A.}~\bibnamefont {Brataas}}, \bibinfo {author} {\bibfnamefont {R.~A.}\
  \bibnamefont {Duine}}, \ and\ \bibinfo {author} {\bibfnamefont
  {M.}~\bibnamefont {Kl\"{a}ui}},\ }\href {\doibase 10.1038/s41586-018-0490-7}
  {\bibfield  {journal} {\bibinfo  {journal} {Nature}\ }\textbf {\bibinfo
  {volume} {561}},\ \bibinfo {pages} {222} (\bibinfo {year}
  {2018})}\BibitemShut {NoStop}%
\bibitem [{\citenamefont {Baltz}\ \emph {et~al.}(2018)\citenamefont {Baltz},
  \citenamefont {Manchon}, \citenamefont {Tsoi}, \citenamefont {Moriyama},
  \citenamefont {Ono},\ and\ \citenamefont {Tserkovnyak}}]{Baltz:RMP2018}%
  \BibitemOpen
  \bibfield  {author} {\bibinfo {author} {\bibfnamefont {V.}~\bibnamefont
  {Baltz}}, \bibinfo {author} {\bibfnamefont {A.}~\bibnamefont {Manchon}},
  \bibinfo {author} {\bibfnamefont {M.}~\bibnamefont {Tsoi}}, \bibinfo {author}
  {\bibfnamefont {T.}~\bibnamefont {Moriyama}}, \bibinfo {author}
  {\bibfnamefont {T.}~\bibnamefont {Ono}}, \ and\ \bibinfo {author}
  {\bibfnamefont {Y.}~\bibnamefont {Tserkovnyak}},\ }\href {\doibase
  10.1103/RevModPhys.90.015005} {\bibfield  {journal} {\bibinfo  {journal}
  {Reviews of Modern Physics}\ }\textbf {\bibinfo {volume} {90}},\ \bibinfo
  {pages} {015005} (\bibinfo {year} {2018})}\BibitemShut {NoStop}%
\bibitem [{\citenamefont {Gomonay}\ \emph {et~al.}(2018)\citenamefont
  {Gomonay}, \citenamefont {Baltz}, \citenamefont {Brataas},\ and\
  \citenamefont {Tserkovnyak}}]{Gomonay:NatPhys2018}%
  \BibitemOpen
  \bibfield  {author} {\bibinfo {author} {\bibfnamefont {O.}~\bibnamefont
  {Gomonay}}, \bibinfo {author} {\bibfnamefont {V.}~\bibnamefont {Baltz}},
  \bibinfo {author} {\bibfnamefont {A.}~\bibnamefont {Brataas}}, \ and\
  \bibinfo {author} {\bibfnamefont {Y.}~\bibnamefont {Tserkovnyak}},\ }\href
  {\doibase 10.1038/s41567-018-0049-4} {\bibfield  {journal} {\bibinfo
  {journal} {Nature Physics}\ }\textbf {\bibinfo {volume} {14}},\ \bibinfo
  {pages} {213} (\bibinfo {year} {2018})}\BibitemShut {NoStop}%
\bibitem [{\citenamefont {Li}\ \emph {et~al.}(2020)\citenamefont {Li},
  \citenamefont {Wilson}, \citenamefont {Cheng}, \citenamefont {Lohmann},
  \citenamefont {Kavand}, \citenamefont {Yuan}, \citenamefont {Aldosary},
  \citenamefont {Agladze}, \citenamefont {Wei}, \citenamefont {Sherwin},\ and\
  \citenamefont {Shi}}]{Li:Nature2020}%
  \BibitemOpen
  \bibfield  {author} {\bibinfo {author} {\bibfnamefont {J.}~\bibnamefont
  {Li}}, \bibinfo {author} {\bibfnamefont {C.~B.}\ \bibnamefont {Wilson}},
  \bibinfo {author} {\bibfnamefont {R.}~\bibnamefont {Cheng}}, \bibinfo
  {author} {\bibfnamefont {M.}~\bibnamefont {Lohmann}}, \bibinfo {author}
  {\bibfnamefont {M.}~\bibnamefont {Kavand}}, \bibinfo {author} {\bibfnamefont
  {W.}~\bibnamefont {Yuan}}, \bibinfo {author} {\bibfnamefont {M.}~\bibnamefont
  {Aldosary}}, \bibinfo {author} {\bibfnamefont {N.}~\bibnamefont {Agladze}},
  \bibinfo {author} {\bibfnamefont {P.}~\bibnamefont {Wei}}, \bibinfo {author}
  {\bibfnamefont {M.~S.}\ \bibnamefont {Sherwin}}, \ and\ \bibinfo {author}
  {\bibfnamefont {J.}~\bibnamefont {Shi}},\ }\href {\doibase
  10.1038/s41586-020-1950-4} {\bibfield  {journal} {\bibinfo  {journal}
  {Nature}\ }\textbf {\bibinfo {volume} {578}},\ \bibinfo {pages} {70}
  (\bibinfo {year} {2020})}\BibitemShut {NoStop}%
\bibitem [{\citenamefont {Vaidya}\ \emph {et~al.}(2020)\citenamefont {Vaidya},
  \citenamefont {Morley}, \citenamefont {van Tol}, \citenamefont {Liu},
  \citenamefont {Cheng}, \citenamefont {Brataas}, \citenamefont {Lederman},\
  and\ \citenamefont {del Barco}}]{Vaidya:Science2020}%
  \BibitemOpen
  \bibfield  {author} {\bibinfo {author} {\bibfnamefont {P.}~\bibnamefont
  {Vaidya}}, \bibinfo {author} {\bibfnamefont {S.~A.}\ \bibnamefont {Morley}},
  \bibinfo {author} {\bibfnamefont {J.}~\bibnamefont {van Tol}}, \bibinfo
  {author} {\bibfnamefont {Y.}~\bibnamefont {Liu}}, \bibinfo {author}
  {\bibfnamefont {R.}~\bibnamefont {Cheng}}, \bibinfo {author} {\bibfnamefont
  {A.}~\bibnamefont {Brataas}}, \bibinfo {author} {\bibfnamefont
  {D.}~\bibnamefont {Lederman}}, \ and\ \bibinfo {author} {\bibfnamefont
  {E.}~\bibnamefont {del Barco}},\ }\href {\doibase 10.1126/science.aaz4247}
  {\bibfield  {journal} {\bibinfo  {journal} {Science}\ }\textbf {\bibinfo
  {volume} {368}},\ \bibinfo {pages} {160} (\bibinfo {year}
  {2020})}\BibitemShut {NoStop}%
\bibitem [{\citenamefont {Wadley}\ \emph {et~al.}(2016)\citenamefont {Wadley},
  \citenamefont {Howells}, \citenamefont {\v{Z}elezn\`{y}}, \citenamefont
  {Andrews}, \citenamefont {Hills}, \citenamefont {Campion}, \citenamefont
  {Novák}, \citenamefont {Olejník}, \citenamefont {Maccherozzi}, \citenamefont
  {Dhesi}, \citenamefont {Martin}, \citenamefont {Wagner}, \citenamefont
  {Wunderlich}, \citenamefont {Freimuth}, \citenamefont {Mokrousov},
  \citenamefont {Kune¨}, \citenamefont {Chauhan}, \citenamefont {Grzybowski},
  \citenamefont {Rushforth}, \citenamefont {Edmonds}, \citenamefont
  {Gallagher},\ and\ \citenamefont {Jungwirth}}]{Wadley:Science2016}%
  \BibitemOpen
  \bibfield  {author} {\bibinfo {author} {\bibfnamefont {P.}~\bibnamefont
  {Wadley}}, \bibinfo {author} {\bibfnamefont {B.}~\bibnamefont {Howells}},
  \bibinfo {author} {\bibfnamefont {J.}~\bibnamefont {\v{Z}elezn\`{y}}},
  \bibinfo {author} {\bibfnamefont {C.}~\bibnamefont {Andrews}}, \bibinfo
  {author} {\bibfnamefont {V.}~\bibnamefont {Hills}}, \bibinfo {author}
  {\bibfnamefont {R.~P.}\ \bibnamefont {Campion}}, \bibinfo {author}
  {\bibfnamefont {V.}~\bibnamefont {Novák}}, \bibinfo {author} {\bibfnamefont
  {K.}~\bibnamefont {Olejník}}, \bibinfo {author} {\bibfnamefont
  {F.}~\bibnamefont {Maccherozzi}}, \bibinfo {author} {\bibfnamefont {S.~S.}\
  \bibnamefont {Dhesi}}, \bibinfo {author} {\bibfnamefont {S.~Y.}\ \bibnamefont
  {Martin}}, \bibinfo {author} {\bibfnamefont {T.}~\bibnamefont {Wagner}},
  \bibinfo {author} {\bibfnamefont {J.}~\bibnamefont {Wunderlich}}, \bibinfo
  {author} {\bibfnamefont {F.}~\bibnamefont {Freimuth}}, \bibinfo {author}
  {\bibfnamefont {Y.}~\bibnamefont {Mokrousov}}, \bibinfo {author}
  {\bibfnamefont {J.}~\bibnamefont {Kune¨}}, \bibinfo {author} {\bibfnamefont
  {J.~S.}\ \bibnamefont {Chauhan}}, \bibinfo {author} {\bibfnamefont {M.~J.}\
  \bibnamefont {Grzybowski}}, \bibinfo {author} {\bibfnamefont {A.~W.}\
  \bibnamefont {Rushforth}}, \bibinfo {author} {\bibfnamefont {K.~W.}\
  \bibnamefont {Edmonds}}, \bibinfo {author} {\bibfnamefont {B.~L.}\
  \bibnamefont {Gallagher}}, \ and\ \bibinfo {author} {\bibfnamefont
  {T.}~\bibnamefont {Jungwirth}},\ }\href
  {http://science.sciencemag.org/sci/early/2016/01/13/science.aab1031.full.pdf}
  {\bibfield  {journal} {\bibinfo  {journal} {Science}\ } (\bibinfo {year}
  {2016})}\BibitemShut {NoStop}%
\bibitem [{\citenamefont {Cheng}\ \emph {et~al.}(2020)\citenamefont {Cheng},
  \citenamefont {Yu}, \citenamefont {Zhu}, \citenamefont {Hwang},\ and\
  \citenamefont {Yang}}]{Cheng:PRL2020}%
  \BibitemOpen
  \bibfield  {author} {\bibinfo {author} {\bibfnamefont {Y.}~\bibnamefont
  {Cheng}}, \bibinfo {author} {\bibfnamefont {S.}~\bibnamefont {Yu}}, \bibinfo
  {author} {\bibfnamefont {M.}~\bibnamefont {Zhu}}, \bibinfo {author}
  {\bibfnamefont {J.}~\bibnamefont {Hwang}}, \ and\ \bibinfo {author}
  {\bibfnamefont {F.}~\bibnamefont {Yang}},\ }\href {\doibase
  10.1103/PhysRevLett.124.027202} {\bibfield  {journal} {\bibinfo  {journal}
  {Physical Review Letters}\ }\textbf {\bibinfo {volume} {124}},\ \bibinfo
  {pages} {027202} (\bibinfo {year} {2020})}\BibitemShut {NoStop}%
\bibitem [{\citenamefont {Cheng}\ \emph {et~al.}(2014)\citenamefont {Cheng},
  \citenamefont {Xiao}, \citenamefont {Niu},\ and\ \citenamefont
  {Brataas}}]{Cheng:PRL2014}%
  \BibitemOpen
  \bibfield  {author} {\bibinfo {author} {\bibfnamefont {R.}~\bibnamefont
  {Cheng}}, \bibinfo {author} {\bibfnamefont {J.}~\bibnamefont {Xiao}},
  \bibinfo {author} {\bibfnamefont {Q.}~\bibnamefont {Niu}}, \ and\ \bibinfo
  {author} {\bibfnamefont {A.}~\bibnamefont {Brataas}},\ }\href
  {http://link.aps.org/doi/10.1103/PhysRevLett.113.057601} {\bibfield
  {journal} {\bibinfo  {journal} {Physical Review Letters}\ }\textbf {\bibinfo
  {volume} {113}},\ \bibinfo {pages} {057601} (\bibinfo {year}
  {2014})}\BibitemShut {NoStop}%
\bibitem [{\citenamefont {Kamra}\ and\ \citenamefont
  {Belzig}(2017)}]{Kamra:PRL2017}%
  \BibitemOpen
  \bibfield  {author} {\bibinfo {author} {\bibfnamefont {A.}~\bibnamefont
  {Kamra}}\ and\ \bibinfo {author} {\bibfnamefont {W.}~\bibnamefont {Belzig}},\
  }\href {https://link.aps.org/doi/10.1103/PhysRevLett.119.197201} {\bibfield
  {journal} {\bibinfo  {journal} {Physical Review Letters}\ }\textbf {\bibinfo
  {volume} {119}},\ \bibinfo {pages} {197201} (\bibinfo {year}
  {2017})}\BibitemShut {NoStop}%
\bibitem [{\citenamefont {Johansen}\ and\ \citenamefont
  {Brataas}(2017)}]{Johansen:PRB2017}%
  \BibitemOpen
  \bibfield  {author} {\bibinfo {author} {\bibfnamefont {O.}~\bibnamefont
  {Johansen}}\ and\ \bibinfo {author} {\bibfnamefont {A.}~\bibnamefont
  {Brataas}},\ }\href {https://link.aps.org/doi/10.1103/PhysRevB.95.220408}
  {\bibfield  {journal} {\bibinfo  {journal} {Physical Review B}\ }\textbf
  {\bibinfo {volume} {95}},\ \bibinfo {pages} {220408} (\bibinfo {year}
  {2017})}\BibitemShut {NoStop}%
\bibitem [{\citenamefont {Brataas}\ \emph {et~al.}(2008)\citenamefont
  {Brataas}, \citenamefont {Tserkovnyak},\ and\ \citenamefont
  {Bauer}}]{Brataas:PRL2008}%
  \BibitemOpen
  \bibfield  {author} {\bibinfo {author} {\bibfnamefont {A.}~\bibnamefont
  {Brataas}}, \bibinfo {author} {\bibfnamefont {Y.}~\bibnamefont
  {Tserkovnyak}}, \ and\ \bibinfo {author} {\bibfnamefont {G.~E.~W.}\
  \bibnamefont {Bauer}},\ }\href
  {http://link.aps.org/abstract/PRL/v101/e037207} {\bibfield  {journal}
  {\bibinfo  {journal} {Physical Review Letters}\ }\textbf {\bibinfo {volume}
  {101}},\ \bibinfo {pages} {037207} (\bibinfo {year} {2008})}\BibitemShut
  {NoStop}%
\bibitem [{\citenamefont {Starikov}\ \emph {et~al.}(2010)\citenamefont
  {Starikov}, \citenamefont {Kelly}, \citenamefont {Brataas}, \citenamefont
  {Tserkovnyak},\ and\ \citenamefont {Bauer}}]{Starikov:PRL2010}%
  \BibitemOpen
  \bibfield  {author} {\bibinfo {author} {\bibfnamefont {A.~A.}\ \bibnamefont
  {Starikov}}, \bibinfo {author} {\bibfnamefont {P.~J.}\ \bibnamefont {Kelly}},
  \bibinfo {author} {\bibfnamefont {A.}~\bibnamefont {Brataas}}, \bibinfo
  {author} {\bibfnamefont {Y.}~\bibnamefont {Tserkovnyak}}, \ and\ \bibinfo
  {author} {\bibfnamefont {G.~E.~W.}\ \bibnamefont {Bauer}},\ }\href
  {http://link.aps.org/doi/10.1103/PhysRevLett.105.236601} {\bibfield
  {journal} {\bibinfo  {journal} {Physical Review Letters}\ }\textbf {\bibinfo
  {volume} {105}},\ \bibinfo {pages} {236601} (\bibinfo {year}
  {2010})}\BibitemShut {NoStop}%
\bibitem [{\citenamefont {Brataas}\ \emph {et~al.}(2011)\citenamefont
  {Brataas}, \citenamefont {Tserkovnyak},\ and\ \citenamefont
  {Bauer}}]{Brataas:PRB2011}%
  \BibitemOpen
  \bibfield  {author} {\bibinfo {author} {\bibfnamefont {A.}~\bibnamefont
  {Brataas}}, \bibinfo {author} {\bibfnamefont {Y.}~\bibnamefont
  {Tserkovnyak}}, \ and\ \bibinfo {author} {\bibfnamefont {G.~E.~W.}\
  \bibnamefont {Bauer}},\ }\href
  {http://link.aps.org/doi/10.1103/PhysRevB.84.054416} {\bibfield  {journal}
  {\bibinfo  {journal} {Physical Review B}\ }\textbf {\bibinfo {volume} {84}},\
  \bibinfo {pages} {054416} (\bibinfo {year} {2011})}\BibitemShut {NoStop}%
\bibitem [{\citenamefont {Liu}\ \emph {et~al.}(2011)\citenamefont {Liu},
  \citenamefont {Starikov}, \citenamefont {Yuan},\ and\ \citenamefont
  {Kelly}}]{Liu:PRB2011}%
  \BibitemOpen
  \bibfield  {author} {\bibinfo {author} {\bibfnamefont {Y.}~\bibnamefont
  {Liu}}, \bibinfo {author} {\bibfnamefont {A.~A.}\ \bibnamefont {Starikov}},
  \bibinfo {author} {\bibfnamefont {Z.}~\bibnamefont {Yuan}}, \ and\ \bibinfo
  {author} {\bibfnamefont {P.~J.}\ \bibnamefont {Kelly}},\ }\href
  {http://link.aps.org/doi/10.1103/PhysRevB.84.014412} {\bibfield  {journal}
  {\bibinfo  {journal} {Physical Review B}\ }\textbf {\bibinfo {volume} {84}},\
  \bibinfo {pages} {014412} (\bibinfo {year} {2011})}\BibitemShut {NoStop}%
\bibitem [{\citenamefont {Foros}\ \emph {et~al.}(2005)\citenamefont {Foros},
  \citenamefont {Brataas}, \citenamefont {Tserkovnyak},\ and\ \citenamefont
  {Bauer}}]{Foros:PRL2005}%
  \BibitemOpen
  \bibfield  {author} {\bibinfo {author} {\bibfnamefont {J.}~\bibnamefont
  {Foros}}, \bibinfo {author} {\bibfnamefont {A.}~\bibnamefont {Brataas}},
  \bibinfo {author} {\bibfnamefont {Y.}~\bibnamefont {Tserkovnyak}}, \ and\
  \bibinfo {author} {\bibfnamefont {G.~E.~W.}\ \bibnamefont {Bauer}},\ }\href
  {<Go to ISI>://000230275500054} {\bibfield  {journal} {\bibinfo  {journal}
  {Physical Review Letters}\ }\textbf {\bibinfo {volume} {95}},\ \bibinfo
  {pages} {016601} (\bibinfo {year} {2005})}\BibitemShut {NoStop}%
\bibitem [{\citenamefont {Ludwig}\ \emph {et~al.}(2020)\citenamefont {Ludwig},
  \citenamefont {Burmistrov}, \citenamefont {Gefen},\ and\ \citenamefont
  {Shnirman}}]{Ludwig:PRR2020}%
  \BibitemOpen
  \bibfield  {author} {\bibinfo {author} {\bibfnamefont {T.}~\bibnamefont
  {Ludwig}}, \bibinfo {author} {\bibfnamefont {I.~S.}\ \bibnamefont
  {Burmistrov}}, \bibinfo {author} {\bibfnamefont {Y.}~\bibnamefont {Gefen}}, \
  and\ \bibinfo {author} {\bibfnamefont {A.}~\bibnamefont {Shnirman}},\ }\href
  {\doibase 10.1103/PhysRevResearch.2.023221} {\bibfield  {journal} {\bibinfo
  {journal} {Physical Review Research}\ }\textbf {\bibinfo {volume} {2}},\
  \bibinfo {pages} {023221} (\bibinfo {year} {2020})}\BibitemShut {NoStop}%
\bibitem [{\citenamefont {B\"{u}ttiker}\ \emph {et~al.}(1993)\citenamefont
  {B\"{u}ttiker}, \citenamefont {Prêtre},\ and\ \citenamefont
  {Thomas}}]{Buttiker:PRL1993}%
  \BibitemOpen
  \bibfield  {author} {\bibinfo {author} {\bibfnamefont {M.}~\bibnamefont
  {B\"{u}ttiker}}, \bibinfo {author} {\bibfnamefont {A.}~\bibnamefont
  {Prêtre}}, \ and\ \bibinfo {author} {\bibfnamefont {H.}~\bibnamefont
  {Thomas}},\ }\href {http://link.aps.org/abstract/PRL/v70/p4114} {\bibfield
  {journal} {\bibinfo  {journal} {Physical Review Letters}\ }\textbf {\bibinfo
  {volume} {70}},\ \bibinfo {pages} {4114} (\bibinfo {year}
  {1993})}\BibitemShut {NoStop}%
\bibitem [{\citenamefont {Waintal}\ \emph {et~al.}(2000)\citenamefont
  {Waintal}, \citenamefont {Myers}, \citenamefont {Brouwer},\ and\
  \citenamefont {Ralph}}]{Waintal:PRB2000}%
  \BibitemOpen
  \bibfield  {author} {\bibinfo {author} {\bibfnamefont {X.}~\bibnamefont
  {Waintal}}, \bibinfo {author} {\bibfnamefont {E.~B.}\ \bibnamefont {Myers}},
  \bibinfo {author} {\bibfnamefont {P.~W.}\ \bibnamefont {Brouwer}}, \ and\
  \bibinfo {author} {\bibfnamefont {D.~C.}\ \bibnamefont {Ralph}},\ }\href
  {http://link.aps.org/abstract/PRB/v62/p12317} {\bibfield  {journal} {\bibinfo
   {journal} {Physical Review B}\ }\textbf {\bibinfo {volume} {62}},\ \bibinfo
  {pages} {12317} (\bibinfo {year} {2000})}\BibitemShut {NoStop}%
\bibitem [{\citenamefont {Brataas}\ \emph {et~al.}(2000)\citenamefont
  {Brataas}, \citenamefont {Nazarov},\ and\ \citenamefont
  {Bauer}}]{Brataas:PRL2000}%
  \BibitemOpen
  \bibfield  {author} {\bibinfo {author} {\bibfnamefont {A.}~\bibnamefont
  {Brataas}}, \bibinfo {author} {\bibfnamefont {Y.~V.}\ \bibnamefont
  {Nazarov}}, \ and\ \bibinfo {author} {\bibfnamefont {G.~E.~W.}\ \bibnamefont
  {Bauer}},\ }\href {\doibase 10.1103/PhysRevLett.84.2481} {\bibfield
  {journal} {\bibinfo  {journal} {Physical Review Letters}\ }\textbf {\bibinfo
  {volume} {84}},\ \bibinfo {pages} {2481} (\bibinfo {year}
  {2000})}\BibitemShut {NoStop}%
\bibitem [{\citenamefont {Brataas}\ \emph {et~al.}(2001)\citenamefont
  {Brataas}, \citenamefont {Nazarov},\ and\ \citenamefont
  {Bauer}}]{Brataas:EPJB2001}%
  \BibitemOpen
  \bibfield  {author} {\bibinfo {author} {\bibfnamefont {A.}~\bibnamefont
  {Brataas}}, \bibinfo {author} {\bibfnamefont {Y.}~\bibnamefont {Nazarov}}, \
  and\ \bibinfo {author} {\bibfnamefont {G.}~\bibnamefont {Bauer}},\ }\href
  {\doibase 10.1007/PL00011139} {\bibfield  {journal} {\bibinfo  {journal}
  {European Physical Journal B}\ }\textbf {\bibinfo {volume} {22}},\ \bibinfo
  {pages} {99} (\bibinfo {year} {2001})}\BibitemShut {NoStop}%
\bibitem [{\citenamefont {Stiles}\ and\ \citenamefont
  {Zangwill}(2002)}]{Stiles:PRB2002}%
  \BibitemOpen
  \bibfield  {author} {\bibinfo {author} {\bibfnamefont {M.~D.}\ \bibnamefont
  {Stiles}}\ and\ \bibinfo {author} {\bibfnamefont {A.}~\bibnamefont
  {Zangwill}},\ }\href {http://link.aps.org/doi/10.1103/PhysRevB.66.014407}
  {\bibfield  {journal} {\bibinfo  {journal} {Physical Review B}\ }\textbf
  {\bibinfo {volume} {66}},\ \bibinfo {pages} {014407} (\bibinfo {year}
  {2002})}\BibitemShut {NoStop}%
\bibitem [{\citenamefont {Rychkov}\ \emph {et~al.}(2009)\citenamefont
  {Rychkov}, \citenamefont {Borlenghi}, \citenamefont {Jaffres}, \citenamefont
  {Fert},\ and\ \citenamefont {Waintal}}]{Rychkov:PRL2009}%
  \BibitemOpen
  \bibfield  {author} {\bibinfo {author} {\bibfnamefont {V.~S.}\ \bibnamefont
  {Rychkov}}, \bibinfo {author} {\bibfnamefont {S.}~\bibnamefont {Borlenghi}},
  \bibinfo {author} {\bibfnamefont {H.}~\bibnamefont {Jaffres}}, \bibinfo
  {author} {\bibfnamefont {A.}~\bibnamefont {Fert}}, \ and\ \bibinfo {author}
  {\bibfnamefont {X.}~\bibnamefont {Waintal}},\ }\href
  {http://link.aps.org/abstract/PRL/v103/e066602} {\bibfield  {journal}
  {\bibinfo  {journal} {Physical Review Letters}\ }\textbf {\bibinfo {volume}
  {103}},\ \bibinfo {pages} {066602} (\bibinfo {year} {2009})}\BibitemShut
  {NoStop}%
\bibitem [{\citenamefont {Hals}\ \emph {et~al.}(2010)\citenamefont {Hals},
  \citenamefont {Brataas},\ and\ \citenamefont {Tserkovnyak}}]{Hals:EPL2010}%
  \BibitemOpen
  \bibfield  {author} {\bibinfo {author} {\bibfnamefont {K.~M.~D.}\
  \bibnamefont {Hals}}, \bibinfo {author} {\bibfnamefont {A.}~\bibnamefont
  {Brataas}}, \ and\ \bibinfo {author} {\bibfnamefont {Y.}~\bibnamefont
  {Tserkovnyak}},\ }\href {http://dx.doi.org/10.1209/0295-5075/90/47002}
  {\bibfield  {journal} {\bibinfo  {journal} {Epl}\ }\textbf {\bibinfo {volume}
  {90}} (\bibinfo {year} {2010})}\BibitemShut {NoStop}%
\bibitem [{\citenamefont {Brouwer}(1998)}]{Brouwer:PRB1998}%
  \BibitemOpen
  \bibfield  {author} {\bibinfo {author} {\bibfnamefont {P.~W.}\ \bibnamefont
  {Brouwer}},\ }\href {http://link.aps.org/abstract/PRB/v58/pR10135} {\bibfield
   {journal} {\bibinfo  {journal} {Physical Review B}\ }\textbf {\bibinfo
  {volume} {58}},\ \bibinfo {pages} {R10135} (\bibinfo {year}
  {1998})}\BibitemShut {NoStop}%
\bibitem [{\citenamefont {Brataas}\ \emph {et~al.}(2002)\citenamefont
  {Brataas}, \citenamefont {Tserkovnyak}, \citenamefont {Bauer},\ and\
  \citenamefont {Halperin}}]{Brataas:PRB2002}%
  \BibitemOpen
  \bibfield  {author} {\bibinfo {author} {\bibfnamefont {A.}~\bibnamefont
  {Brataas}}, \bibinfo {author} {\bibfnamefont {Y.}~\bibnamefont
  {Tserkovnyak}}, \bibinfo {author} {\bibfnamefont {G.~E.~W.}\ \bibnamefont
  {Bauer}}, \ and\ \bibinfo {author} {\bibfnamefont {B.~I.}\ \bibnamefont
  {Halperin}},\ }\href {http://link.aps.org/doi/10.1103/PhysRevB.66.060404}
  {\bibfield  {journal} {\bibinfo  {journal} {Physical Review B}\ }\textbf
  {\bibinfo {volume} {66}},\ \bibinfo {pages} {060404} (\bibinfo {year}
  {2002})}\BibitemShut {NoStop}%
\bibitem [{\citenamefont {Tserkovnyak}\ \emph {et~al.}(2005)\citenamefont
  {Tserkovnyak}, \citenamefont {Brataas}, \citenamefont {Bauer},\ and\
  \citenamefont {Halperin}}]{Tserkovnyak:RMP2005}%
  \BibitemOpen
  \bibfield  {author} {\bibinfo {author} {\bibfnamefont {Y.}~\bibnamefont
  {Tserkovnyak}}, \bibinfo {author} {\bibfnamefont {A.}~\bibnamefont
  {Brataas}}, \bibinfo {author} {\bibfnamefont {G.~E.~W.}\ \bibnamefont
  {Bauer}}, \ and\ \bibinfo {author} {\bibfnamefont {B.~I.}\ \bibnamefont
  {Halperin}},\ }\href {http://link.aps.org/abstract/RMP/v77/p1375} {\bibfield
  {journal} {\bibinfo  {journal} {Reviews of Modern Physics}\ }\textbf
  {\bibinfo {volume} {77}},\ \bibinfo {pages} {1375} (\bibinfo {year}
  {2005})}\BibitemShut {NoStop}%
\bibitem [{\citenamefont {Moskalets}\ and\ \citenamefont
  {B\"{u}ttiker}(2004)}]{Moskalets:PRB2004}%
  \BibitemOpen
  \bibfield  {author} {\bibinfo {author} {\bibfnamefont {M.}~\bibnamefont
  {Moskalets}}\ and\ \bibinfo {author} {\bibfnamefont {M.}~\bibnamefont
  {B\"{u}ttiker}},\ }\href {http://link.aps.org/abstract/PRB/v70/e245305}
  {\bibfield  {journal} {\bibinfo  {journal} {Physical Review B (Condensed
  Matter and Materials Physics)}\ }\textbf {\bibinfo {volume} {70}},\ \bibinfo
  {pages} {245305} (\bibinfo {year} {2004})}\BibitemShut {NoStop}%
\bibitem [{\citenamefont {Brataas}\ \emph {et~al.}(2006)\citenamefont
  {Brataas}, \citenamefont {Bauer},\ and\ \citenamefont
  {Kelly}}]{Brataas:PhysRep2006}%
  \BibitemOpen
  \bibfield  {author} {\bibinfo {author} {\bibfnamefont {A.}~\bibnamefont
  {Brataas}}, \bibinfo {author} {\bibfnamefont {G.~E.~W.}\ \bibnamefont
  {Bauer}}, \ and\ \bibinfo {author} {\bibfnamefont {P.~J.}\ \bibnamefont
  {Kelly}},\ }\href
  {http://www.sciencedirect.com/science/article/B6TVP-4JF97B4-1/2/499d943549f8b52e7a2fdde6e9fdc472}
  {\bibfield  {journal} {\bibinfo  {journal} {Physics Reports}\ }\textbf
  {\bibinfo {volume} {427}},\ \bibinfo {pages} {157} (\bibinfo {year}
  {2006})}\BibitemShut {NoStop}%
\bibitem [{\citenamefont {Bender}\ \emph {et~al.}(2019)\citenamefont {Bender},
  \citenamefont {Kamra}, \citenamefont {Belzig},\ and\ \citenamefont
  {Duine}}]{Bender:PRL2019}%
  \BibitemOpen
  \bibfield  {author} {\bibinfo {author} {\bibfnamefont {S.~A.}\ \bibnamefont
  {Bender}}, \bibinfo {author} {\bibfnamefont {A.}~\bibnamefont {Kamra}},
  \bibinfo {author} {\bibfnamefont {W.}~\bibnamefont {Belzig}}, \ and\ \bibinfo
  {author} {\bibfnamefont {R.~A.}\ \bibnamefont {Duine}},\ }\href {\doibase
  10.1103/PhysRevLett.122.187701} {\bibfield  {journal} {\bibinfo  {journal}
  {Physical Review Letters}\ }\textbf {\bibinfo {volume} {122}},\ \bibinfo
  {pages} {187701} (\bibinfo {year} {2019})}\BibitemShut {NoStop}%
\bibitem [{\citenamefont {Nakayama}\ \emph {et~al.}(2013)\citenamefont
  {Nakayama}, \citenamefont {Althammer}, \citenamefont {Chen}, \citenamefont
  {Uchida}, \citenamefont {Kajiwara}, \citenamefont {Kikuchi}, \citenamefont
  {Ohtani}, \citenamefont {GeprŠgs}, \citenamefont {Opel}, \citenamefont
  {Takahashi}, \citenamefont {Gross}, \citenamefont {Bauer}, \citenamefont
  {Goennenwein},\ and\ \citenamefont {Saitoh}}]{Nakayama:PRL2013}%
  \BibitemOpen
  \bibfield  {author} {\bibinfo {author} {\bibfnamefont {H.}~\bibnamefont
  {Nakayama}}, \bibinfo {author} {\bibfnamefont {M.}~\bibnamefont {Althammer}},
  \bibinfo {author} {\bibfnamefont {Y.~T.}\ \bibnamefont {Chen}}, \bibinfo
  {author} {\bibfnamefont {K.}~\bibnamefont {Uchida}}, \bibinfo {author}
  {\bibfnamefont {Y.}~\bibnamefont {Kajiwara}}, \bibinfo {author}
  {\bibfnamefont {D.}~\bibnamefont {Kikuchi}}, \bibinfo {author} {\bibfnamefont
  {T.}~\bibnamefont {Ohtani}}, \bibinfo {author} {\bibfnamefont
  {S.}~\bibnamefont {GeprŠgs}}, \bibinfo {author} {\bibfnamefont
  {M.}~\bibnamefont {Opel}}, \bibinfo {author} {\bibfnamefont {S.}~\bibnamefont
  {Takahashi}}, \bibinfo {author} {\bibfnamefont {R.}~\bibnamefont {Gross}},
  \bibinfo {author} {\bibfnamefont {G.~E.~W.}\ \bibnamefont {Bauer}}, \bibinfo
  {author} {\bibfnamefont {S.~T.~B.}\ \bibnamefont {Goennenwein}}, \ and\
  \bibinfo {author} {\bibfnamefont {E.}~\bibnamefont {Saitoh}},\ }\href
  {http://link.aps.org/doi/10.1103/PhysRevLett.110.206601} {\bibfield
  {journal} {\bibinfo  {journal} {Physical Review Letters}\ }\textbf {\bibinfo
  {volume} {110}},\ \bibinfo {pages} {206601} (\bibinfo {year}
  {2013})}\BibitemShut {NoStop}%
\bibitem [{\citenamefont {Schep}\ \emph {et~al.}(1997)\citenamefont {Schep},
  \citenamefont {van Hoof}, \citenamefont {Kelly}, \citenamefont {Bauer},\ and\
  \citenamefont {Inglesfield}}]{Schep:PRB1997}%
  \BibitemOpen
  \bibfield  {author} {\bibinfo {author} {\bibfnamefont {K.~M.}\ \bibnamefont
  {Schep}}, \bibinfo {author} {\bibfnamefont {J.~B. A.~N.}\ \bibnamefont {van
  Hoof}}, \bibinfo {author} {\bibfnamefont {P.~J.}\ \bibnamefont {Kelly}},
  \bibinfo {author} {\bibfnamefont {G.~E.~W.}\ \bibnamefont {Bauer}}, \ and\
  \bibinfo {author} {\bibfnamefont {J.~E.}\ \bibnamefont {Inglesfield}},\
  }\href {\doibase 10.1103/PhysRevB.56.10805} {\bibfield  {journal} {\bibinfo
  {journal} {Physical Review B}\ }\textbf {\bibinfo {volume} {56}},\ \bibinfo
  {pages} {10805} (\bibinfo {year} {1997})}\BibitemShut {NoStop}%
\bibitem [{\citenamefont {Xia}\ \emph {et~al.}(2002)\citenamefont {Xia},
  \citenamefont {Kelly}, \citenamefont {Bauer}, \citenamefont {Brataas},\ and\
  \citenamefont {Turek}}]{Xia:PRB2002}%
  \BibitemOpen
  \bibfield  {author} {\bibinfo {author} {\bibfnamefont {K.}~\bibnamefont
  {Xia}}, \bibinfo {author} {\bibfnamefont {P.~J.}\ \bibnamefont {Kelly}},
  \bibinfo {author} {\bibfnamefont {G.~E.~W.}\ \bibnamefont {Bauer}}, \bibinfo
  {author} {\bibfnamefont {A.}~\bibnamefont {Brataas}}, \ and\ \bibinfo
  {author} {\bibfnamefont {I.}~\bibnamefont {Turek}},\ }\href
  {http://link.aps.org/doi/10.1103/PhysRevB.65.220401} {\bibfield  {journal}
  {\bibinfo  {journal} {Physical Review B}\ }\textbf {\bibinfo {volume} {65}},\
  \bibinfo {pages} {220401} (\bibinfo {year} {2002})}\BibitemShut {NoStop}%
\bibitem [{\citenamefont {Zwierzycki}\ \emph {et~al.}(2005)\citenamefont
  {Zwierzycki}, \citenamefont {Tserkovnyak}, \citenamefont {Kelly},
  \citenamefont {Brataas},\ and\ \citenamefont {Bauer}}]{Zwierzycki:PRB2005}%
  \BibitemOpen
  \bibfield  {author} {\bibinfo {author} {\bibfnamefont {M.}~\bibnamefont
  {Zwierzycki}}, \bibinfo {author} {\bibfnamefont {Y.}~\bibnamefont
  {Tserkovnyak}}, \bibinfo {author} {\bibfnamefont {P.~J.}\ \bibnamefont
  {Kelly}}, \bibinfo {author} {\bibfnamefont {A.}~\bibnamefont {Brataas}}, \
  and\ \bibinfo {author} {\bibfnamefont {G.~E.~W.}\ \bibnamefont {Bauer}},\
  }\href {http://link.aps.org/doi/10.1103/PhysRevB.71.064420} {\bibfield
  {journal} {\bibinfo  {journal} {Physical Review B}\ }\textbf {\bibinfo
  {volume} {71}},\ \bibinfo {pages} {064420} (\bibinfo {year}
  {2005})}\BibitemShut {NoStop}%
\bibitem [{\citenamefont {Ciccarelli}\ \emph {et~al.}(2015)\citenamefont
  {Ciccarelli}, \citenamefont {Hals}, \citenamefont {Irvine}, \citenamefont
  {Novak}, \citenamefont {Tserkovnyak}, \citenamefont {Kurebayashi},
  \citenamefont {Brataas},\ and\ \citenamefont
  {Ferguson}}]{Ciccarelli:NatNan2015}%
  \BibitemOpen
  \bibfield  {author} {\bibinfo {author} {\bibfnamefont {C.}~\bibnamefont
  {Ciccarelli}}, \bibinfo {author} {\bibfnamefont {K.~M.~D.}\ \bibnamefont
  {Hals}}, \bibinfo {author} {\bibfnamefont {A.}~\bibnamefont {Irvine}},
  \bibinfo {author} {\bibfnamefont {V.}~\bibnamefont {Novak}}, \bibinfo
  {author} {\bibfnamefont {Y.}~\bibnamefont {Tserkovnyak}}, \bibinfo {author}
  {\bibfnamefont {H.}~\bibnamefont {Kurebayashi}}, \bibinfo {author}
  {\bibfnamefont {A.}~\bibnamefont {Brataas}}, \ and\ \bibinfo {author}
  {\bibfnamefont {A.}~\bibnamefont {Ferguson}},\ }\href {\doibase
  10.1038/nnano.2014.252
  http://www.nature.com/nnano/journal/v10/n1/abs/nnano.2014.252.html#supplementary-information}
  {\bibfield  {journal} {\bibinfo  {journal} {Nature Nanotechnology}\ }\textbf
  {\bibinfo {volume} {10}},\ \bibinfo {pages} {50} (\bibinfo {year}
  {2015})}\BibitemShut {NoStop}%
\bibitem [{\citenamefont {Rammer}\ and\ \citenamefont
  {Smith}(1986)}]{Rammer:RMP1986}%
  \BibitemOpen
  \bibfield  {author} {\bibinfo {author} {\bibfnamefont {J.}~\bibnamefont
  {Rammer}}\ and\ \bibinfo {author} {\bibfnamefont {H.}~\bibnamefont {Smith}},\
  }\href {http://link.aps.org/abstract/RMP/v58/p323} {\bibfield  {journal}
  {\bibinfo  {journal} {Reviews of Modern Physics}\ }\textbf {\bibinfo {volume}
  {58}},\ \bibinfo {pages} {323} (\bibinfo {year} {1986})}\BibitemShut
  {NoStop}%
\end{thebibliography}%

\end{document}